# The Extraction of Community Structures from Publication Networks to Support Ethnographic Observations of Field Differences in Scientific Communication

Theresa Velden*, Carl Lagoze
tvelden@umich.edu, clagoze@umich.edu
School of Information, University of Michigan, 4365 North Quad, Ann Arbor 48109-1285, MI, USA

**Abstract**

The scientific community of researchers in a research specialty is an important unit of analysis for understanding the field specific shaping of scientific communication practices. These scientific communities are, however, a challenging unit of analysis to capture and compare because they overlap, have fuzzy boundaries, and evolve over time. We describe a network analytic approach that reveals the complexities of these communities through examination of their publication networks in combination with insights from ethnographic field studies. We suggest that the structures revealed indicate overlapping sub-communities within a research specialty and we provide evidence that they differ in disciplinary orientation and research practices. By mapping the community structures of scientific fields we aim to increase confidence about the domain of validity of ethnographic observations as well as of collaborative patterns extracted from publication networks thereby enabling the systematic study of field differences. The network analytic methods presented include methods to optimize the delineation of a bibliographic data set in order to adequately represent a research specialty, and methods to extract community structures from this data. We demonstrate the application of these methods in a case study of two research specialties in the physical and chemical sciences.

*: *Corresponding author*

## 1 Introduction

This article introduces a method for the analysis of scientific publication networks that supports ethnographic observations of field differences in scientific communication. This work is part of a larger research program to develop a mixed method approach that combines qualitative-ethnographic and quantitative-network analytic methods to systematically study and develop theoretical explanations for differences of scientific communication practices across research specialties. The emergence of the World Wide Web and adoption of web based technologies to support scientific communication has revived research interest into field differences in scientific communication practices (Kling & Kim 2000, Fry & Talja 2007). However, comparative studies of scientific fields with an ethnographic depth of understanding of research practices and social behaviors are rare (Gläser 2006), and bibliometric studies provide only narrow insights into field specific practices because of the limitations of publication data to reflect underlying research and communication processes (Lievrouw 1990). The current scarcity of empirical material comparing scientific communities at the field level inhibits theory development to answer fundamental questions such as how field-specific characteristics of scientific knowledge interact with the social ordering of scientific communities (Gläser 2006), and how epistemic cultures shape communication practices (Cronin 2003, Fry & Talja 2004) and attitudes towards openness and sharing (Velden 2011a).

To advance research on these questions, we propose to bridge the methodological gap with an integrated approach that combines the analysis of publication networks with ethnographic field studies. Comparative studies of scientific communities at field-level face several challenges that this approach is designed to overcome. Whereas bibliographic data and the social networks constructed from them are popular objects of study in bibliometrics (e.g. Jansen et al. 2010, Lambiotte & Panzaras 2009, Guimera et al. 2005, Melin & Persson 1996, Kretschmer 1994), quantitative structural and statistical analyses suffer from a lack of explanatory power with regard to the underlying processes and their meaning to the actors involved (Gläser & Laudel 2001, Lievrouw 1990). Ethnography on the other hand is well suited for generating understandings of actors, their practices and motivations, but difficult to scale. This suggests combining the two approaches (Lievrouw 1990) that hitherto have been rarely used in a truly synergistic manner. However, how local observations scale-up to and tie in with behavioral patterns at field-level is not obvious: how do behavioral characteristics at the field level emerge from local practices, and what is the domain of validity for observations made?

The challenges posed by these open questions are exacerbated by the complexity of scientific fields as units of analysis. The salience of the ***research specialty***[1] as a self-organizing unit of researchers that is responsible for the production of new scientific knowledge has been recognized by the sociology of science in the 1960's and 1970's (Morris & Van der Veer Martens2008, Becher & Towler 2001). Those studies also highlighted the overlapping, layered configuration of research specialties, their fuzzy boundaries, and their dynamic temporal evolution. This structural complexity and temporal fluidity provides challenges for the delineation of research specialties as analytic units for comparative studies across fields (Zitt 2005), and for the valid extrapolation of local ethnographic observations of individuals and groups to the collective level of scientific communities within a research specialty.

Rather than simply aim for the delineation of a research specialty, we suggest the development of methods that help us navigate the complexity of layered scientific communities that make up a research specialty and transcend its boundaries. In this article we report first results using network analytic methods to reveal community structures within scientific fields, thereby contributing to the literature about the mapping of research specialties recently reviewed by Morris & Van der Veer Martens (2008). Our aim is to identify scientific communities as units of analysis for comparative studies of the influence of material and epistemic characteristics of research on collaboration and communication behaviors. The identification of such units of analysis will support the scaling-up of ethnographies and the extraction of typical collaboration patterns from publication networks for meaningful comparisons of research specialties.

This article is organized as follows: section 2 provides a review of related literature and a summary of our overall research strategy that uses a mixed-method approach to study field differences in scientific communication. The network analytic methods used in this article are described in section 3. They include methods to extract community structures from bibliographic data, and methods to support field delineation, that is the definition of an adequate bibliographic data set to extract community structures from. The extraction of community structures is achieved through the quantitative analysis of topical and social affinities between topic areas within a research specialty (section 3.4), and the visualization of community structures by geographic and topical overlay maps on field-wide group collaboration networks (section 3.5). The methods supporting the definition and preparation of the bibliographic data set consist of an approach to monitor the quality of author name disambiguation and the resulting reduction in network distortion due to name homonymy (section 4.1), and a set of heuristic methods to optimize field delineation by controlling recall and precision in the lexical query (section 4.2). We demonstrate in section 5 the application of these methods by reporting results from a case study of two fields in the physical and chemical sciences, and discuss implications of the empirical results for the methodological approach in section 6.

## 2 Background

In this section we review related work that contextualizes and motivates our efforts to develop a mixed method approach to study field differences in scientific communication, and we provide an overview of the approach as developed so far.

### 2.1 Related Work

We focus on three central themes in reviewing the related literature: efforts to generate a theoretical understanding of how epistemic differences between research specialties interact with the social organization of scientific communities, leading to field differences in communication practices; the challenges of capturing and describing scientific communities as unit of analysis in comparative studies; and motivations and strategies for a mixed method approach that combines network analysis with ethnographic field studies.

The emergence of the World Wide Web has triggered a renewed and intensified research interest in differences between research specialties and their communication practices, and how such differences explain variation in the form and speed of take-up of new technologies to support knowledge sharing in scientific communities (e.g. Walsh & Bayma 1996, Kling & Kim 2000, Birnholtz & Bietz 2003, Cronin 2003, Kling et al. 2003, Kling et al. 2004, Nentwich 2005, Fry 2007, Hine 2008, Cana 2010). These scientific inquiries face substantial theoretical challenges, such as the open question of how to conceptualize the dynamics of change in complex socio-technical systems (discussed e.g. by Geels 2007), and the lack of theoretical understanding of differences in the intellectual and social organization of research specialties and how these differences shape scientific communication practices. What are the salient characteristics of research specialties that influence communication practices, and what is the appropriate level of analysis (disciplines, sub-disciplines, research specialties) for meaningful comparison and contextualization for communication practices observed?

To describe how the intellectual and the social organization of a research community are intertwined, Knorr Cetina (1999) has developed the notion of 'epistemic cultures', a concept that has become popular in the information science literature concerned with differences in scientific communication practices across fields (e.g. VanHouse 2002, Cronin 2003, Beaulieu et al. 2007, Baus 2010). However, Knorr Cetina's ethnographic case study of high energy physics and molecular biology derives salient cultural features bottom-up, delivering only loosely coupled descriptions of the two epistemic cultures (Gläser 2006). The concept of 'epistemic cultures' does not provide an analytical framework of the scientific research process that would support a systematic comparison between fields (ibid).

In a more systematic approach to capture the socio-intellectual characteristics of intellectual fields Whitley (2000) introduces a taxonomy that distinguishes disciplines by degree of practical and strategic dependence, and by degree of task and strategic uncertainty. Empirical studies have used the taxonomy to explain field specific practices of data sharing, the creation of web-based information and communication resources, and the use of open access repositories (Birnholtz & Bietz 2003, Fry&Talja 2007, Cana 2010). Problematic for the application of Whitley's taxonomy however is the determination of the appropriate unit of analysis. Fry and Talja (2007) observe that Whitley has developed his taxonomy at the level of well-established disciplines (e.g. '20th century physics'), and that it falls short of capturing more dynamic, possibly transdisciplinary research collectives.

Another related problem is the conceptualization of scientific fields as reputational work organizations (Gläser 2006). The organizational sociology perspective applied by Whitley provides only a partial insight into the role of scientific communication in the process of knowledge production. Therefore, the link between the taxonomic classification of a research field and how that classification explains field specific communication practices remains vague. Empirical studies in information science and bibliometrics oftentimes fall back on coarse disciplinary or sub-disciplinary divisions for field comparisons. However, smaller, more specialized entities within (sub)disciplines expose significant differences in social organization and research culture (Becher & Towler 2001, Galison 1997, Mulkay 1977), and it has been suggested that an appropriate level of analysis for the comparison of field specific communication behaviors has to be sought at a finer level of granularity, such as the research specialty (e.g. Fry & Talja 2004, Jamali & Nicholas 2008).

Theoretical support for this view is provided by Gläser's theory for the collective production of scientific knowledge in research specialties. The theory is derived from a thorough review of empirical studies in science studies, and it ascribes an indirect but crucial coordinating role to the shared knowledge base of the ***scientific community***[2] in a research specialty (Gläser 2006, p.18). By the orientation of the self-directed, autonomous research activities of community members toward contributing to this shared knowledge base a social ordering emerges that enables the collective knowledge production process (ibid). This model implies that scientific communication plays a key role in the coordination (social ordering) of the collective production of knowledge within a research specialty. Gläser hypothesizes that differences in the social ordering of communities can be linked to differences in the nature of their respective knowledge bases. We would expect such differences to be reflected also in the scientific communication practices of a community.

Unfortunately, research specialties are difficult to capture as a unit of analysis, last but not least due to the self-similarity of science (Zitt 2005). The boundaries of research specialties are fuzzy, as even field members do not necessarily agree on where to draw the line since research specialties overlap and evolve in time. Bibliometric field delineation, that is the extraction of a subset of publications from a bibliographic database to represent either all the authors or all the relevant publications in a research field is a difficult task, with no simple generic solution available.

Journal-level field delineation that rests on the selection of core journals in a field has been shown to be problematic due to the wide scope of most journals and the wide spread of specialized literature across many different journals (Huang et al. 2011, Aksnes et al. 1999). Instead, lexical queries that retrieve individual publications by matching query terms to their bibliographic metadata support a more precise retrieval of relevant literature. However, they tend to suffer from a lack of recall, especially for newly emerging and interdisciplinary fields (Mogoutov & Kahane 2007). Sophisticated hybrid approaches have been developed to delineate and monitor such fields. They start from a high precision and low recall seed of publications generated by a lexical query, and extend this seed either through citation coupling (Laurens et al. 2010, Zitt 2006) or keyword based retrieval (Mogoutov & Kahane 2007). The quality of the field delineation is further affected by the coverage of databases available to the analyst, the extent of access to that database, and the quality and granularity of the data. Depending on the scientific domain, different databases or a combination of databases provide better results (Strotmann & Zhao 2010).

In this study we retrieve our data from the Web of Science database (WoS) by Thomson Reuters) since it has reasonable coverage in established fields in physics and chemistry (Moed 2005), especially for the 20-year time range we were interested in. Due to economic restrictions in access to the complete WoS, we could not use the hybrid approach developed by Zitt & Bassecoulard (2006) that requires access to the 'cited by' links from a cited publication to the citing publication, a restriction pointed out by Huang et al. (2011) and Mogoutov & Kahane (2007). Instead we start from a lexical query, and then use heuristic network analytic methods to improve recall and precision. The new methods that we introduce in section 4.2 for the optimization of the lexical query are unique due to their transparency with regard to the social topology of the field, as discussed in 6.1.

We suggest a mixed method approach will be essential to produce the empirical base critically needed to develop a theoretical understanding of the field specific shaping of communication practices. By combining ethnography with the network analysis of bibliographic data we can deliver nuanced and situated understandings of communication practices along with insights into community structures and behavioral patterns at the collective level, as suggested by (Lievrouw 1990). These structural patterns indicate the domain of validity for local observations, and further support the quantification of some aspects of field differences, like collaboration intensity or group sizes and structures.

Strategically, our motivation for combining ethnography with network analysis has strong parallels with the advantages of a 'network ethnography' approach advocated by Howard (2002)[3]. He perceives of the combination of ethnographic fieldwork with network analysis as a synergistic, transdisciplinary method with distinct advantages for the study of communication in modern organizations over new media. Aiming to develop more rigorous methods to obtain generalizable qualitative data, he values this particular method combination because it improves the construction and strategic sampling of field sites for ethnographic research.

However, border crossings that combine ethnographic studies with quantitative network analysis are rare in the study of research communities. There exists a strong disciplinary divide between the quantitative focus of fields such as bibliometrics, scientometrics and the science of networks[4], and the predominance of qualitatively oriented work in fields such as the sociology of science and science and technology studies (Gläser&Laudel 2007, Gläser&Laudel 2001, Van den Besselaar 2000). Some of the very few exceptions are Geiger & Ribes (2011) who develop the idea of 'trace ethnography' to study the use of IT based systems in cyberscience, Cambrosio et al. (22004) who analyze large-scale collaborative research in an area of bio-medicine by making use of heterogeneous networks that link entities such as humans, laboratories and molecules, and Zuccala (2004) who explores invisible college structures in a field of mathematics extending work by Lievrouw et al. (1987) and Crane (1972) who used mixed method approaches to study invisible colleges. This article contributes to this latter strand of work by revealing community structures within scientific fields with the specific aim to support the scaling-up of ethnographic analysis for comparative studies of scientific communication practices.

### 2.2 A Mixed Method Approach to Study Field Differences

Our approach to the comparative study of scientific communities can be characterized as 'integration' of network analysis and ethnographic studies, following the terminology by Moran-Ellis (2006) to distinguish and characterize mixed method approaches. This means we use both methods in an interdependent way such that results obtained by one method feed directly into the other.

Figure 1 depicts schematically the interdependency of important steps in the research process, showing how insights generated with one method feed into the other method, and vice versa. Our starting point has been ethnographic field studies of five research groups from two research specialties in the physical and chemical sciences. One of the authors visited each group for several weeks, conducting observations and semi-structured interviews with most of the members of each research group (step 1). Based on the understanding of the research field and local research practices gained during these visits and supported by feedback from knowledgeable study participants, we developed a lexical query for retrieval of data from a bibliographic database to capture the publication output of the field (step 2). Throughout this process, observations from the field studies together with network analytic data inform decisions on how to adapt and optimize the lexical query, a process that is described in detail in section 4.2.1. and section 4.2.2. below. The co-author network is generated from the publication data retrieved, and various network features are extracted and visualized. These include the structural composition of co-author clusters and between-cluster interlinking patterns. The interpretation of network features in terms of real-life group organization patterns and collaboration scenarios is derived from interviews and joint examination of network visualizations with senior researchers at the field sites (step 3). The network features are quantified for comparisons across research specialties (Velden et al. 2010), generating further research questions. Finally, community structures within the fields are extracted (step 4), as detailed in section 3.4 and section 3.5. Observations from the field studies support the interpretation of those structures that in return contextualize some of the observations and may guide the strategic sampling of further field sites or interview partners.

Eventually, findings are synthesized to identify field specific patterns of communication and to analyze epistemic, material, and social factors shaping the underlying practices (step 5). Preliminary findings on field differences in openness and sharing behaviors in two groups are reported in Velden (2011a, 2013). The insights into community structures reported in this article will guide the refinement of future research into field specific communication practices.

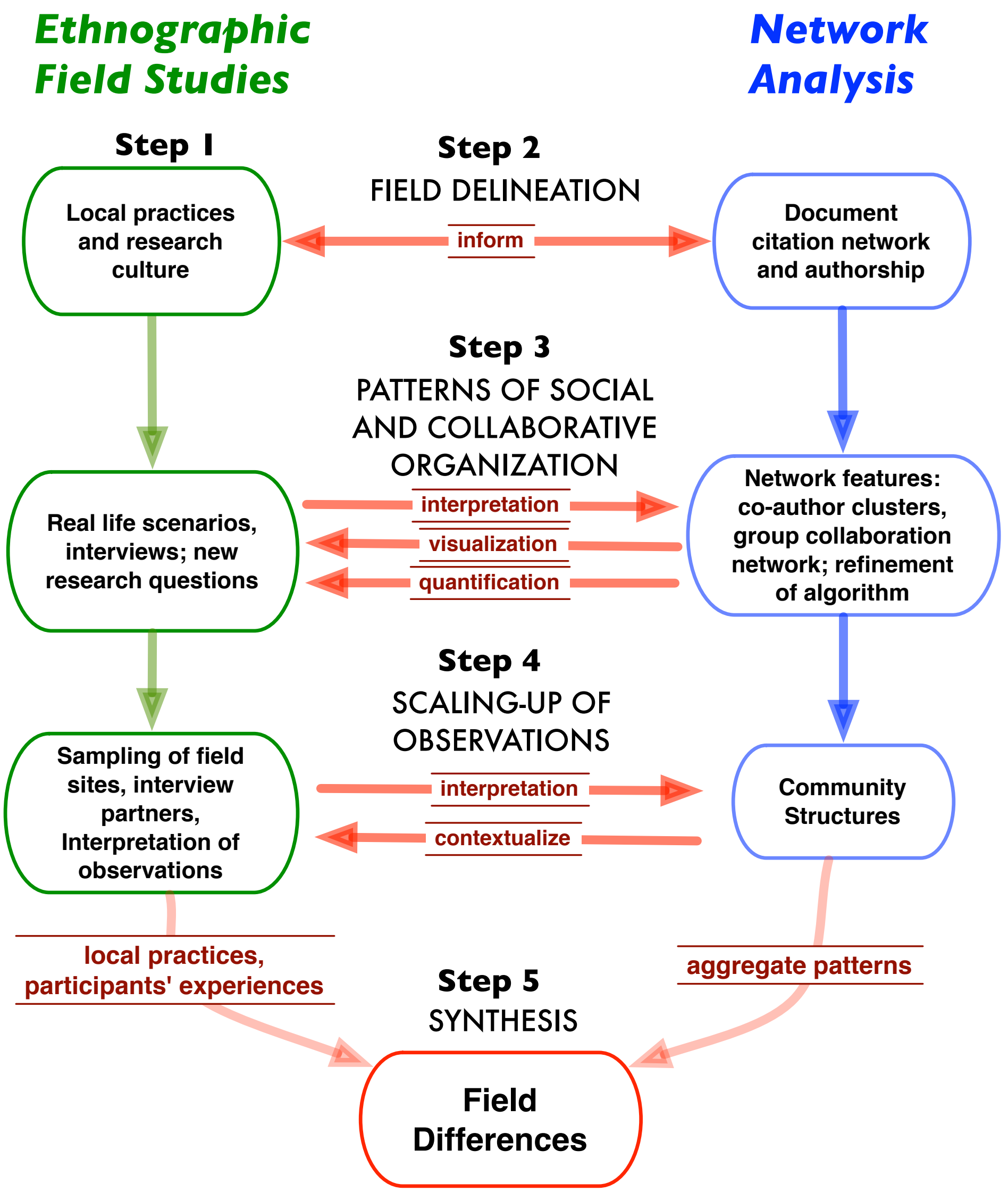

**Figure 1: Schematic representation of the combination of qualitative and quantitative methods in a mixed method approach to study and compare social practices in scientific communities across scientific fields.**

The significance of this combined ethnographic and network-analytic approach is that it allows us on the one hand to develop in-depth contextual understandings of researchers' behaviors through ethnographic observation and analysis. The ethnographic component produces nuanced insights into processes underlying the abstracted patterns of co-author and citation networks, making actors' motivations and every-day practices accessible to analysis and interpretation. The network analysis of the publication data, on the other hand, provides us with quantifiable, aggregate data on community-wide behaviors to the extent that they are signified by citation or co-authorship. This way we can capture emergent community-wide patterns of group organization, collaboration, and knowledge integration, which are only partially visible from qualitative, locally specific observations.

Beyond providing complementary data, the integration of the two methods is a dynamic process and ameliorates shortcomings of either method in the following ways: The data visualizations generate new questions such as what is the role of small co-author clusters in a field that sustain their publication activity over the entire time period? Simultaneously, the ethnographic analysis of research processes and social organization critically supports the interpretation of the network features. It thereby ensures that quantifiable, structural differences between fields can be linked to specific behavioral patterns, and it further guides meaningful refinements of the network analysis (e.g. the distinction between transfer and collaboration links between co-author clusters). Further, the network analytic extraction of community structures reported in this article supports the scaling-up of meaningful units of analysis. By making community structures visible, it supports the strategic sampling of research sites and increases confidence in the validity of ethnographic observations and their extrapolation to larger collective aggregates at the community level.

## 3 Methods

The focus in this article is on the development of network analytic methods to reveal community structures within a scientific field that can then feed into the mixed methods approach described above. We construct two basic network types from bibliographic data of the publication output in a scientific field, the undirected co-author network and the directed document citation network. The nodes in the co-author network represent authors and links represent joint co-authorship of a publication. Node size indicates the number of publications by the author in the data set and link strength indicates the number of publications in the data set that two authors have co-authored together. In the document citation network nodes represent publications, and a directed link represents a reference made from one publication to another. Figure 2 provides an overview of the main steps of processing the two networks, indicating how the network constructs retrieved are used in the analysis. The following sections describe in more detail the basic network analytic constructs used in this study.

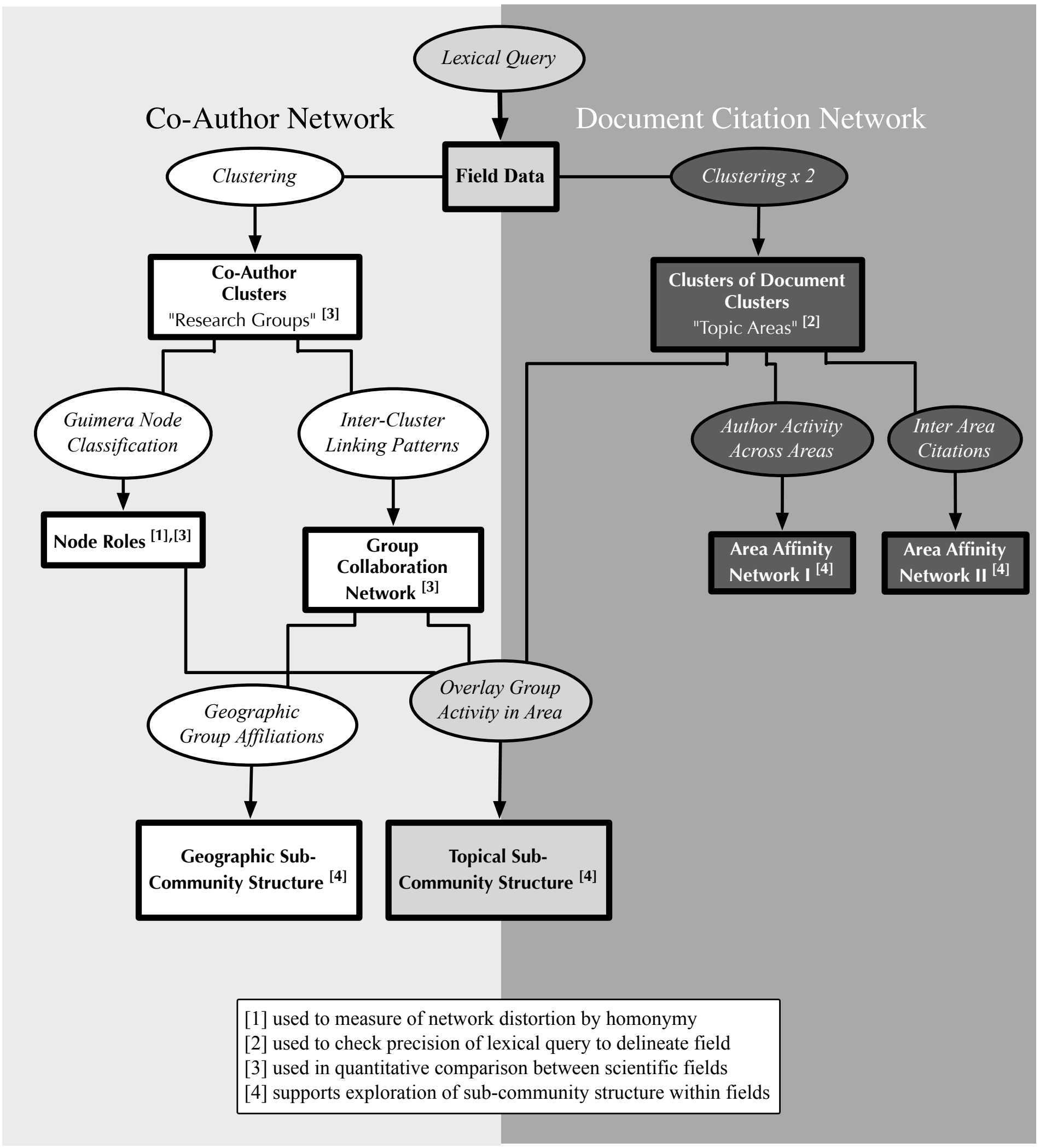

**Figure 2: Overview on network analysis steps. Details on network clustering are given in section 3.2.1, on the Guimera node role classification in section 3.2.3, and on the group collaboration network with geographical and topic area overlays in section 3.2.5. Details on the creation of topic area affinity networks are provided in section 3.2.4. The way we use node role classification to assess network distortion is explained in section 4.1. The way we use topic areas to support field delineation is described in section 4.2.2. Results on the area affinity networks are reported in section 5.2. The quantitative comparison of the two fields with regard to the size of group collaboration networks and results on the extraction of sub-community structure of the two fields are reported in section 5.3.**

### 3.1 Network Clustering

In a first step each network is clustered to extract its modular substructure. The clustering algorithm used is Rosvall's information theoretic algorithm that models information flows (Rosvall & Bergstrom 2008). It has shown superior behavior in retrieving clusters from benchmark networks with clusters of a range of different sizes compared to other popular clustering algorithms (Lancichinetti & Fortunato 2009). However, any clustering method based on the optimization of a global measure will probably have an intrinsic resolution limit (Lancichinetti & Fortunato 2009) such that the clusters retrieved need to be considered with caution. However our mixed method approach mediates this caution. Judging from the insights we got in our ethnographic field studies about the identity of and relationships between groups, we have found that the co-author clusters that this algorithm retrieves for the co-author networks included in this study are very plausible representations of those social collectives that act as primary contributors of original work in a field. For example, the algorithm retrieved from our data set for field 2 a large experimental physics group (82 co-authors during 20 years), as well as a much smaller theoretical physics group (16 co-authors). In the other hand, in an example from field 1, two chemistry groups led by two PIs that are friends and have been working very closely over an extended period of time in this particular field were retrieved as a single cluster of co-authors. Further, by manual lookup of homepages on the World Wide Web we have found that many small co-author clusters[5] in field 2 correspond to theory groups. This observation also suggests that the clustering algorithm is capable to retrieve socially relevant units from the network. Hence, the clusters retrieved can be identified as ***functional research groups***[6] (Seglen & Asknes 2000), either research groups led by a single ***Principal Investigator***[7](PI) or closely collaborating teams of research groups with several PIs (Velden et al. 2010).

Using Rosvall's clustering algorithm twice on the citation network results in a high concentration of documents in several large clusters, such that the clusters of minimum size of 1-2% of all documents capture around 80-90% of documents in the field data set. We take these document clusters to represent topic areas within the research specialty[8]. In this study we select 2% as minimum size for topic areas and focus our analysis on those major topic areas within each field.

### 3.2 Reference Inclusion Rate

The degree to which publications within one of the topic areas within the field cite work within or outside the field could potentially be meaningful to assess the integration and overlap of that topic area with the field. By the reference inclusion rate of a topic area we refer to the proportion of references from publications published within that topic area that point to previously published publications in the entire field data set. We suggest that calculating the reference inclusion rates of a topic area can be a useful measure of how comprehensively a topic area is represented by the field data set. Assuming that the data set captures a chosen field well, a lower reference inclusion rate could then be interpreted as indication that a topic area overlaps with another field or several other fields, whereas topic areas with higher reference inclusion rates would seem to be more strongly contained within the field. Specifically, to track the temporal evolution of reference inclusion rates, we used a moving 5-year window, calculating for each year between 1996-2010 the reference inclusion rate considering only references to publications in the previous 5 years.

### 3.3 Node Role Classification

To capture structural differences between modular networks Guimera et al. (2007) introduce a classification of nodes in clustered networks that is derived from their linking patterns. At the top level this classification distinguishes between hub nodes and non-hub nodes. Hub nodes have disproportionally many cluster internal links relative to the average inside-cluster degree of the nodes in the respective cluster, whereas non-hubs have below average inside cluster links. Based on their outside links to nodes in other clusters, hubs are further sub-divided into 'provincial hubs', 'connector nodes', and 'satellite connector nodes' where the former have least outside links, and the latter have links to many other clusters. Similarly, non-hubs are subdivided into 'ultra-peripheral nodes', 'peripheral nodes', 'connector nodes', and 'satellite connector nodes'.

We apply this classification to the nodes in the clustered co-author network. The stratification of nodes in the co-author network by their node role can then be used to detect network distortions due to author name ambiguity, as explained in more detail in section 4.1.

### 3.4 Topic Area Association and Affinity Networks

To explore patterns of affinity or antagonism between topic areas we quantify the connectivity between topic areas in terms of citations and in terms of author activity overlap, i.e. share of publications of authors who contribute to several topic areas. For the null model of expected connectivity we assume that an author randomly selects a document in another topic area to cite, or randomly selects another topic area to publish in. According to this null model, the probability of a citation from an article in one area (source area) to an article in another area (target area) is proportional to the relative proportion of numbers of articles in the target area compared to the other major topic areas that a citation might point to. Similarly, the null model probability that an author publishing in a topic area (source area) choses for his or her out-of-area publications a specific topic area (target area) is proportional to the relative size of that target area compared to the other potential target areas[9].

We quantify the association between areas as the empirical (actual) deviation from this null model. We treat the null model as providing the expected values for a multinominal distribution. We apply a chi square goodness of fit test to evaluate whether for a specific source area the actual distribution of out-of-area citations or out-of-area publications across potential target areas deviates from the hypothesized null model distribution. The residuals quantify for each target area the positive or negative deviation from the expectation values. They are given by:

$$\text{residuals} = (\text{count_actual} - \text{count_expected}) / \text{count_expected}$$

where the expected counts are calculated according to the null model as:

$$\text{count_expected} := \text{relative target area size} * \text{count_total}$$

where

*relative target area size*: proportional size of target area relative to all potential target areas

*count_total (citations):* sum over all publications in the source area of the sum of all their citations to any publication in any of the target areas

*count_total (authors):* sum over all authors in the source area of the sum of all their publications in any of the target areas

We interpret positive residuals as affinities, pointing to a positive and disproportionally strong association between two topic areas. To visualize the affinities between areas we create a directed weighted network where nodes represent topic areas and the strength of links between nodes is given by those positive residuals. Note that zero affinity does not mean the absence of any links between two areas, but could either represent the assumed background of area connectivity that scales with area sizes or a negative deviation from this assumed background (antagonism).

### 3.5 Group Collaboration Network Overlays

In Velden et al. (2010) we derive a conceptual and algorithmic distinction between ’transfer’ links and ’group collaboration’ links for a co-author network. The latter linking patterns are understood as bibliometric traces of intensive collaborative relationships between research groups in contrast to ’transfer links’ that indicate ephemeral exchange or transfer processes, such as group-to-group career migration of individuals or one-off, service-type collaborations. Applying this distinction we extract from the clustered co-author network a group collaboration network where nodes represent co-author clusters and links represent actual inter-group collaboration. These collaboration networks represent the collaborative core of a field. To study community substructures in a field we overlay those collaboration networks with geographic affiliation information or topical affiliation information on the groups as described below and illustrated e.g. by figure 10 in the results section.

The geographic affiliation of groups is extracted from the institutional affiliations listed for most publications in the WoS database. We evaluate this information only at the continent level. Each cluster is represented by all the publications co-authored by at least one of its authors. For this set of publications we determine the country affiliation that is most often listed. In case the second placed country is listed at least 50% as many times as the most often listed country, and if these two countries belong to different continents, we assign a mixed, two continent geographical affiliation.

The topical overlay maps show for each topic area how active the groups in the collaboration network are in the respective areas. To increase precision, we only consider the publications of hub nodes in the cluster[10]; if a cluster has no hub nodes we consider the publication output of the entire collective of authors in the cluster. A group's activity in a topic area is calculated as the proportion of its publications that is part of the document cluster representing that topic area. We use node color to represent group activity in the topic overlay on the group collaboration network.

## 4 Data

The bibliographic data used in this study has been selected to represent two scientific fields: a research specialty in synthetic chemistry (field 1), and an experimentally oriented research specialty that overlaps with chemistry and physics (field 2). Field 1 is defined by a chemical reaction mechanism, the creation of molecular catalysts that enable this type of reactions, and the application of these reactions in chemical syntheses. Field 2 is driven by solving fundamental conceptual questions, and defined by a physically defined class of objects that provide opportunities to study these fundamental questions. Both fields are dominated by basic research conducted at Universities and research institutes rather than applied research. These fields were selected in the context of our mixed method research into field differences in communication behaviors.

We retrieve the bibliographic data from the WoS for a time period of twenty years (1991- 2010). We use a lexical query developed to identify the publication output of the two fields, an iterative process of field delineation described in section 4.2. Since authors are identified in the WoS database by their last name and initials only, name homonymy[11] is problematic, distorting the co-author networks that can be constructed. To improve the accuracy of the network we apply an author disambiguation algorithm, described below in section 4.1.

### 4.1 Author Name Disambiguation

To address the problem of name homonymy we apply an algorithm, which was developed and evaluated in Velden et al. (2011), to disambiguate author names. We use descriptive network statistics to monitor the improvement made in resolving the undistorted network structure, a method that avoids the expensive investment in creating a representative ground truth sample (Velden et al. 2011). This approach to measuring network distortion rests on the assumption that, given perfect resolution of authors' identities, the structural position of an author in the network and the commonality of his or her last name are uncorrelated. Therefore, an uneven distribution of common last names across the seven classes of node roles in the co-author network (introduced in section 3.2) signifies network distortion due to name ambiguity.

The cumulative probability distributions of last name commonalities by node role for all nodes in the giant component of the co-author network are shown before and after disambiguation in figure 3. Successful disambiguation is reflected by close agreement between the curves for the different node roles. Most curves still reflect a long tailed distribution of last name commonalities, but without the very heavy tails the curves showed before disambiguation. Very common names have become more equally distributed between node roles and disambiguation has been effective (though not perfect) for both data sets.

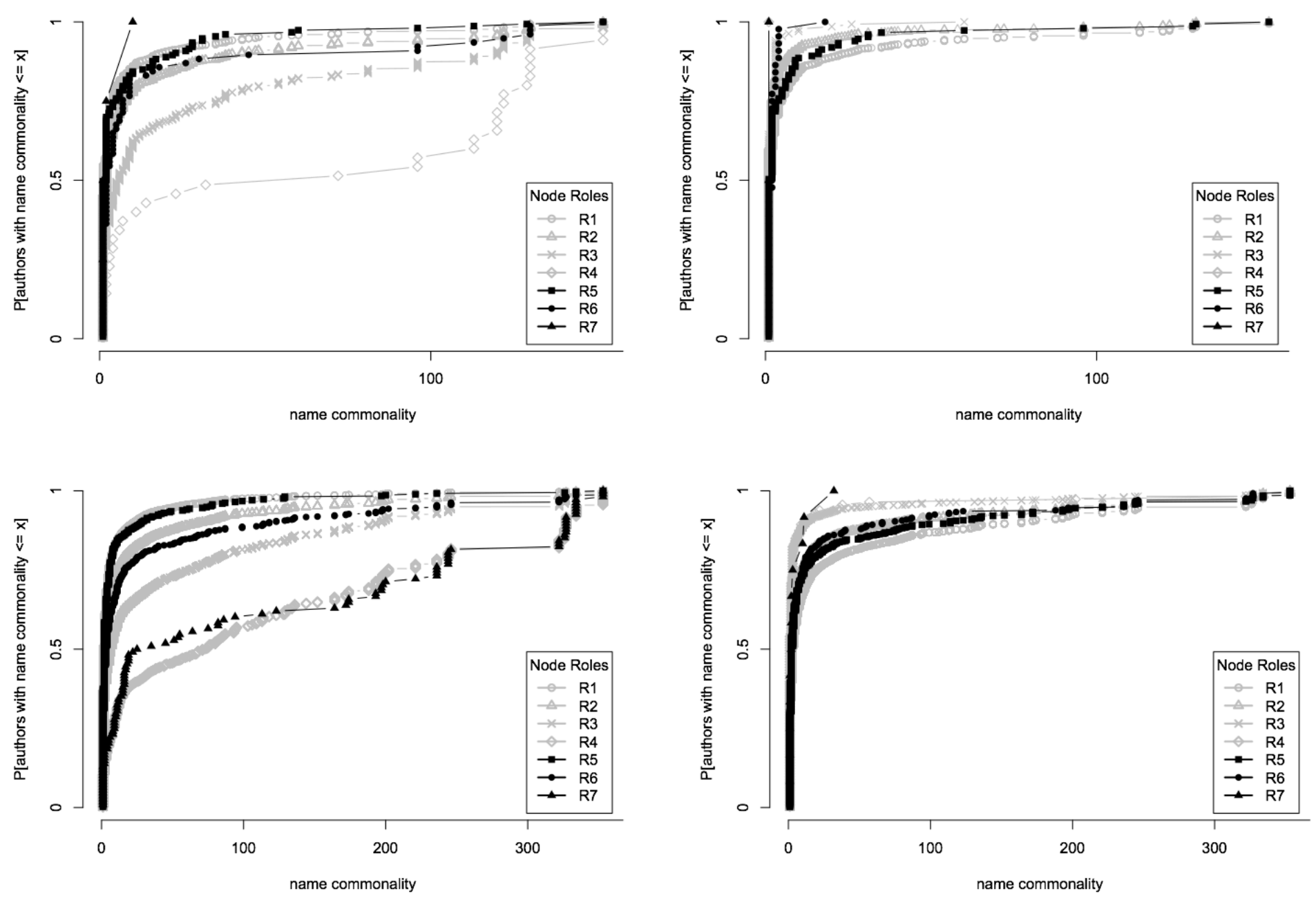

**Figure 3: Cumulative probability distributions of name commonalities for field 1 (top) and field 2 (bottom) before disambiguation (left) and after disambiguation (right).**

The effect of the disambiguation on the number of unique authors identified in the data set is shown in table 1. As could be expected we obtain a higher number of authors after disambiguation, and the number of authors with only one publication increases substantially. However, since we filter out 1-time authors from the data to obtain more concise co-author networks (Velden et al. 2010), the net result in terms of total number of authors in the co-author networks is roughly neutral.

| | **Field 1** | | **Field 2** | |
|---|---|---|---|---|
| | non-disambiguated | disambiguated | non-disambiguated | disambiguated |
| # of publications | 14,599 | | 65,003 | |
| # of authors | 25,500 | 27,946 | 105,776 | 131,285 |
| # of 1-time authors | 15,937 | 18,664 | 64,869 | 90,006 |

Table 1: Data set before and after author name disambiguation

### 4.2 Field Delineation

We start with a standard approach to publication-level field delineation by using a lexical query to retrieve all publications during a defined time frame that have specific terms in the title, abstract, or keyword field of the bibliographic record. In addition we filter the records retrieved by subject categories offered by Web of Science[12].

The task of developing the lexical query of terms characteristic for each research field is rather unproblematic for field 1 since this field is defined by work on a class of catalyzed chemical reactions that is known by a specific name that is highly standardized and commonly used in titles or abstracts of publications in this field. However, for field 2 the field delineation process is challenging and we had to go through several iterations. We checked the progress made by a combination of participant feedback and network analyses. In the next two sections we introduce two heuristic network analytic methods to guide the optimization of the lexical query, one for checking recall, section 4.2.1., and one for checking precision, section 4.2.2.

#### 4.2.1 Recall of Lexical Query

The recall of the lexical query - whether one is capturing all of the relevant publications - can be tracked by checking whether the publications of renowned scientists in the research field are actually included in the data set retrieved.

To this end, one needs to carefully select well-known researchers that span the breadth of the field. For several individuals we retrieve all their publications as indexed by the WoS database. These individuals are the leaders of the research groups included in the ethnographic field study as well as selected researchers who were designated by field study participants as being important figures in the field.

Since the research interests of prolific authors are often spread over several research specialties, we need to identify the subset of publications of a researcher that is relevant to the research field of interest. Hellsten et al. (2007) found that document clusters in the self-citation network of an author's entire work distinguish the research topics that a researcher has worked on. Hence, we construct for each selected researcher the directed, unweighted self-citation network of his or her publication output. Using a clustering algorithm[13] by Rosvall & Bergstrom (2008) we extract document clusters to identify research topics within a researcher's work. An example of such a clustered self-citation network is provided in figure 4.

We verified the appropriateness of the clustering provided by this algorithm by reviewing the resulting clustered network of their publications with the research group leaders of the five groups in our ethnographic field study. Each of them felt it was returning a comprehensive representation of their research interests, as exemplified by the following quote:

*I think that if you look at this, that's my life. Yeah, I think you've got virtually everything. Any subarea (missing?)… no, not really. Nope.* [PI, field 1]

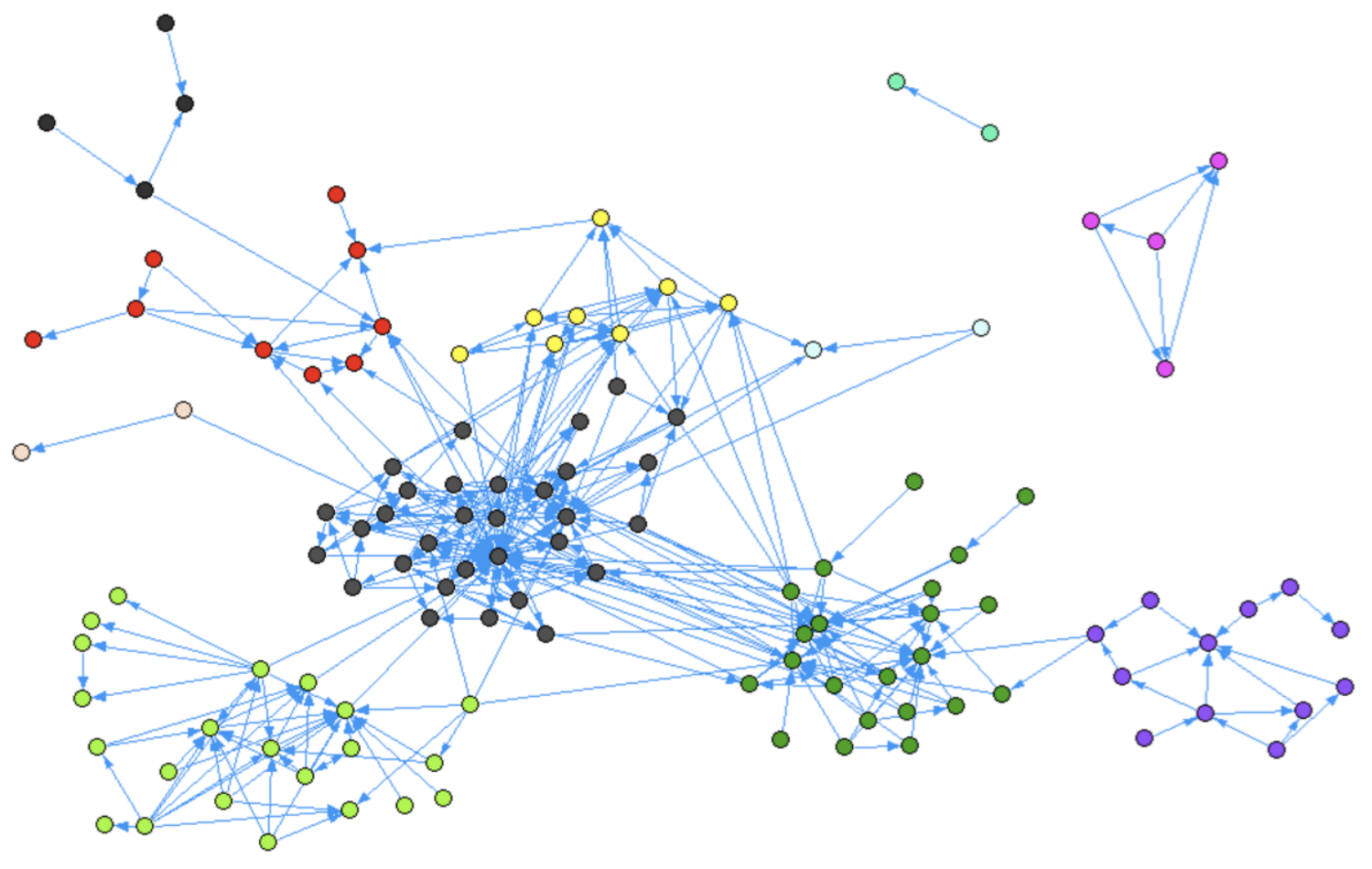

**Figure 4: Selfcitation network of a research group leader. Nodes are documents, node colors indicate cluster membership (interpreted as research topics), and links between documents**

**represent citations between documents (self-citations of the author).**

Next we need to identify those document clusters from a researcher's self-citation network that pertain to the targeted research field. To this end we manually inspect publication titles, and assign a tentative topic description to each cluster. These decision whether a cluster belongs to the research specialty or not seems relatively unproblematic at least for those clusters of publications that can be considered core contributions of that researcher to the research specialty. We then can calculate the overlap between the field data and those clusters. Whenever such a cluster is only weakly represented we inspect titles and abstracts to search for plausible, sufficiently standardized terms to add to the lexical query to improve recall.

#### 4.2.2 Precision of Lexical Query

To check precision of the lexical query means to check whether it captures irrelevant publications that belong to another research field with no or minimal overlap with the scientific field of interest. Since research specialties are overlapping and scientists oftentimes work in several research specialties it is not always straightforward to delineate a research field based on topical distinctions alone. To control for extensions into neighboring fields we investigate how well the various topic areas within a field (retrieved from clustering the citation network, see section 3.2.1.) connect to one another in terms of authors contributing to several of them.

Figure 5 shows two heat maps to visualize and compare inter-area connectivity for two variants of the lexical query for field 2. It depicts the relative participation of authors across the four largest topic areas for an early version of the lexical query and the final version of the query that generated the data set used in this study. The matrix values represent the residuals of the expected association of topic areas derived from author activity, introduced in 3.2.4. To the left the resulting affinity networks are shown, highlighting positive associations between topic areas. The heat map for the historic lexical query for field 2 shows a checkerboard pattern that indicates that the four largest topic areas form two almost disjoint sets of areas, as shown also by the affinity network. For the final lexical query for field 2 the heat map pattern indicates that although the two largest areas 1 and 3 are antagonistic, areas 1, 2, and 4 are connected and a crucial connection between those three areas and area 3 is made by area 3 authors contributing disproportionally also to area 2, and to a lesser extent, also to area 4, as shown in the affinity network. Because this author activity analysis for the historical query indicated that two of the four largest areas had minimal overlap with the other two largest areas, the question arose whether this disconnect signaled the unintended inclusion of another research field through the lexical query.

Upon inspection of the titles of publications in area 2 and 4 of the network for the historical query, we found that they were mostly associated with a term referring to particles of a dimension that went beyond the focus of those studied in the research specialty initially identified. A participant whose research stretched both areas had suggested it in the course of the field study. These particles are produced with chemical methods and require a quite different skill set, and it looked like few researchers had ventured (yet?) to bridge those areas.

The evidence of social discontinuity depicted in the heat map of the intermediary query prompted us to reconsider and to opt against inclusion of this more chemically oriented research direction in the data set for field 2. This kind of judgment requires a deeper understanding of the distinctions made by study participants and highlights the value of a mixed method approach.

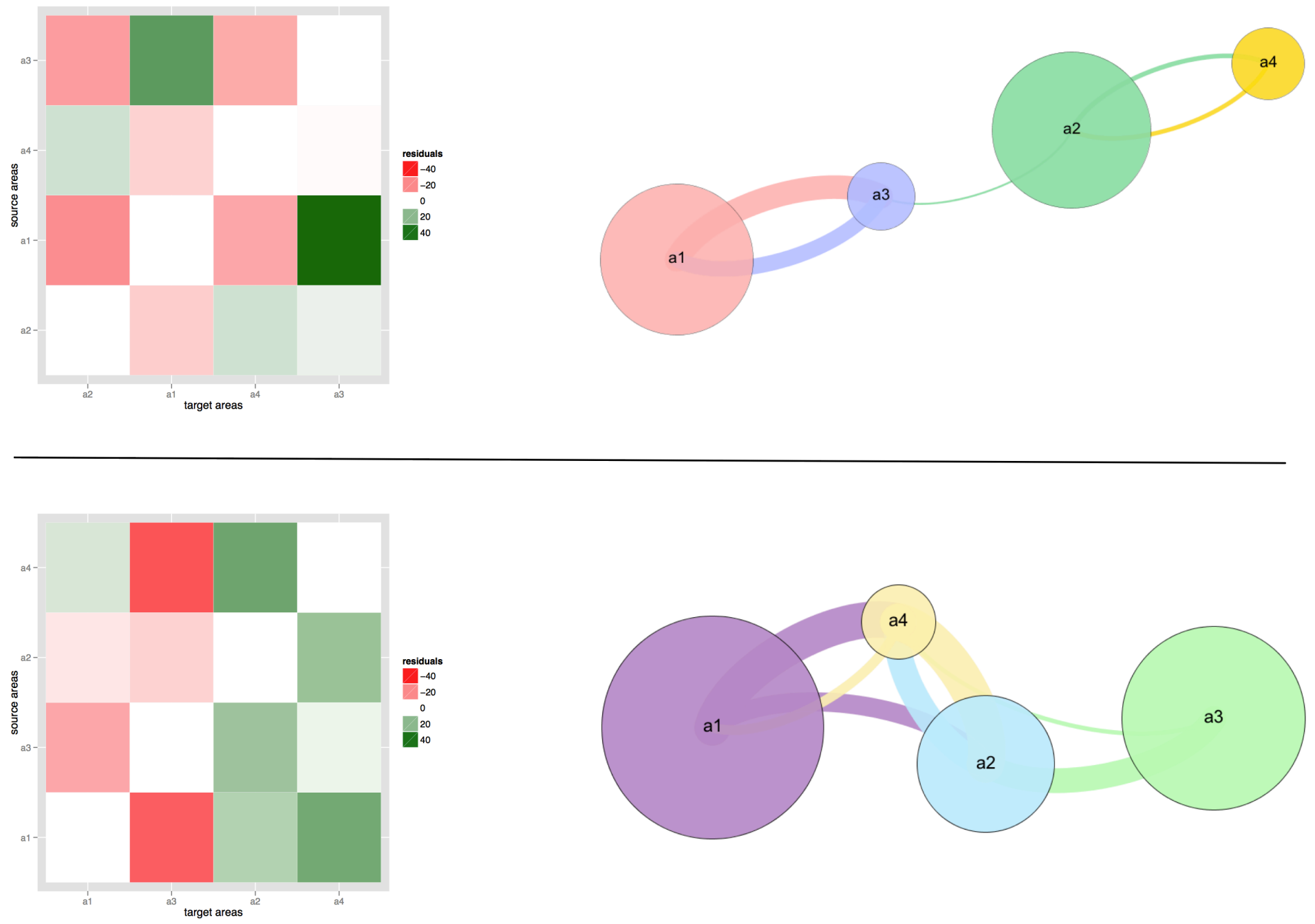

**Figure 5: Heat map and affinity network of author activity based area associations between the largest four topic areas retrieved by two different versions of the lexical query for field 2, and early one (left) and the final one (right). Heatmap colors indicate strength of positive (green) or negative (red) associations between a source area and a target area. Link colors indicate source node of directed links.**

## 5 Extraction of Community Structures Using Network Analysis

Basic properties of the co-author networks and the document citation networks constructed from the bibliographic data for fields 1 and 2 are given in table 2. In the supplementary material we show the literature growth in fields 1 and 2 between 1991 and 2010. Field 1 has a distinct s-shaped growth curve with a strong growth over a ten year time period, more than quadrupling the annual publication output between 1996-2005. The growth curve in field 2 shows a weaker s-shape, just about doubling its annual publication output in the ten-year period between 1995-2004.

| | **Field 1** | **Field 2** |
|---|---|---|
| **Document Citation Network** | | |
| # of documents excl. singletons | 13,910 | 55,648 |
| # of document clusters | 509 | 2,704 |
| # of clusters of document clusters | 102 | 1,069 |
| **Co-author Network** | | |
| # authors after filtering | 9,116 | 40,808 |
| # of clusters | 1,132 | 4,270 |
| # of nodes in giant (proportion) | 6,645 (72.9%) | 33,203 (81.4%) |
| # of clusters in giant (proportion) | 532 (47.0%) | 2,086 (48.6%) |
| Average cluster size in giant (median) | 19.7 (11) | 27.5 (13) |
| # of linked clusters in collaboration network (proportion) | 48 (9.0%) | 477 (22.9%) |

Table 2: Network Sizes and Properties

In the following we report on the extraction of community structures from the document citation networks (section 5.1), and combine them with group collaboration networks to visualize community structures in geographic and topical overlay maps (section 5.2).

**5.1 Topic Areas Derived from Citation Network**

For field 1 we obtain 6 major topic areas that fulfill the 2% size criterion (see section 3.2.1). Together they represent 96.1% of documents in the document citation network. For field 2 we obtain 11 major topic areas that represent 84.9% of documents in the document citation network. Figure 6 shows for each field the network of citations between the major topic areas. We find that field 1 is dominated by one large topic area that is more than twice as large as the other five topic areas taken together. The area sizes in field 2 are more balanced, with three rather large topic areas and five mid-size topic areas, and three small topic areas.

**5.1.1 Topic Area Labeling**

We manually assign topic descriptions to the topic areas extracted from the citation network based on a combination of resources: first we familiarize ourselves with research topics within the field through ethnographic observation and interviews with researchers in the field as well as the reading of review articles of the field. Next we extract and browse the publication metadata for each of the document sets in a cluster, specifically the most frequent author names, the titles of articles, as well as the titles of the most frequent journals published in[14]. Further, we use string searches on text files with the bibliographic data for the publications in each document cluster to double check the plausibility of assignments made by counting the prevalence of technical terms that have a certain topic specific relevance and discriminatory power. The tables depicted in figures 9 and 10 provide an overview on the resulting topic assignments for the major document clusters in each field. We find that frequently topic areas indicate disciplinary orientations, along with specific research foci. For example, in field 1 there are three topic areas associated with polymer chemistry, however each with a different research theme.

**5.1.2 Topic Area Evolution**

Next we look at the growth of topic areas over time, as well as the reference inclusion rates for each topic area (defined in section 3.2.2.). The results are shown in figure 6. We make the following observations:

Growth in field 1 is driven by the dramatic growth of topic area *1:Catalyst Development and Organic Chemistry*. This observation conforms to popular narratives about the history of the field. Accordingly, it experienced a significant breakthrough in the late 1980's and early 1990' when a few pioneers succeeded to produce a specific type of molecular catalysts that turned out to have great application potential for syntheses in organic chemistry. The research group leader of an organic chemistry group in our field study describes how he got interested in this new type of reactions and catalysts very early in the development of the field:

> *"So, during the last years this [reaction] has been the main research area [of the group] and it still is the most important area in terms of deployment of PhD students and co-workers […] When we started it was not obvious what one could do with it and the catalysts were not there yet. There existed isolated works and it was simply the personal assessment, that this is something that could be very useful for organic chemistry. That's why we started that. […] That was, as I already mentioned, at the beginning of the 90's. At this time no broadly useable catalysts existed. At the time when we started, there existed* *no well-defined catalyst that had a high tolerance for functional groups. [...] the reason to do first studies [was] motivated by the application, applicability in natural product synthesis. So, that was the beginning."*

This group has become one of the major contributors to the field, active both in catalyst development and applications of those catalysts in organic synthesis. The major driving force for the strong growth of field 1 after the mid 1990's has been the availability of a couple of effective catalysts and their application in organic synthesis.

Further, we note that in field 1 area *3:Organometallics & Applications in Inorganic Chemistry* has a particularly low reference inclusion rate, setting it apart form the other areas.

The growth dynamics in field 2 are more complex as four major topic areas (*1:Physical Chemistry & Chemical Physics*, *2:Surface Science*, *3:Materials Science*, and *9:Inorganic Chemistry*) show significant growth and share the majority of publication output. Relative growth is most pronounced for areas 2 and 3. The three research groups included in our ethnographic field study were active primarily in area 1 and area 5, hence we heard accounts about the growth of research areas and associated community dynamics based on experiences of researchers active in those two topic areas. Work in topic area 1 has a tradition reaching back into the 1980's and was fairly established by the beginning of the 1990's with almost 300 articles per annum in our data set (figure 6). It has seen significant growth in the 1990's, as research groups in the community has increasingly made use of synchrotron radiation to study geometric and electronic properties of free clusters. Those involved in research in topic area 5 focus on dynamic processes when clusters are submitted to high-intensity laser light and pulsed laser light. Based on our data set this area is much smaller than area 1, and its size is stagnating. However, 'on the ground' excitement at the end of the first decade in the 2000's is great due to a new generation of radiation facilities (x-ray free electron lasers, or XFELs). A handful of these expensive facilities are under construction worldwide and coming online. At this time the user base is still very small and access very competitive, but all of the three groups in our field study had been involved in first pioneering experiments at these new facilities. This suggests that this area of research within field 2 is viable and further evolving.

The reference inclusion rates are highest for topic areas *2:Surface Science*, *4:Physical Chemistry & Chemical Physics*, and *5:Dynamic Processes*, and lowest for areas *7:Inorganic Chemistry*, *8:Catalysis*, *9:Inorganic Chemistry*, and *11:Bio-Inorganic Chemistry*.

In the next two sections we investigate whether we can detect positive associations or affinities between topic areas based on their topical relatedness or their social connectedness (see section 3.2.4. for definitions).

#### 5.1.3 Topical Relatedness between Topic Areas

An affinity between two topic areas in terms of citation indicates some form of topical relatedness beyond a common background of topical connectivity present in the field. The respective affinity networks for field 1 and field 2 are depicted in figure 7, and the full data tables of positive or negative association values (residuals) are provided in the supplementary material. We find the following patterns of topical relatedness for the two fields summarized in the affinity diagrams:

In field 1, area *4:Polymer Chemistry/Organometallics* and area *2:Polymer Chemistry* stand out as receiving disproportionally many citations from other areas. Area 1, that dominates all other topic areas in size, disproportionally often cites those two areas, and detailed inspection of the document clusters in area 1 reveals that some polymer chemistry document clusters have been subsumed into area 1, offering

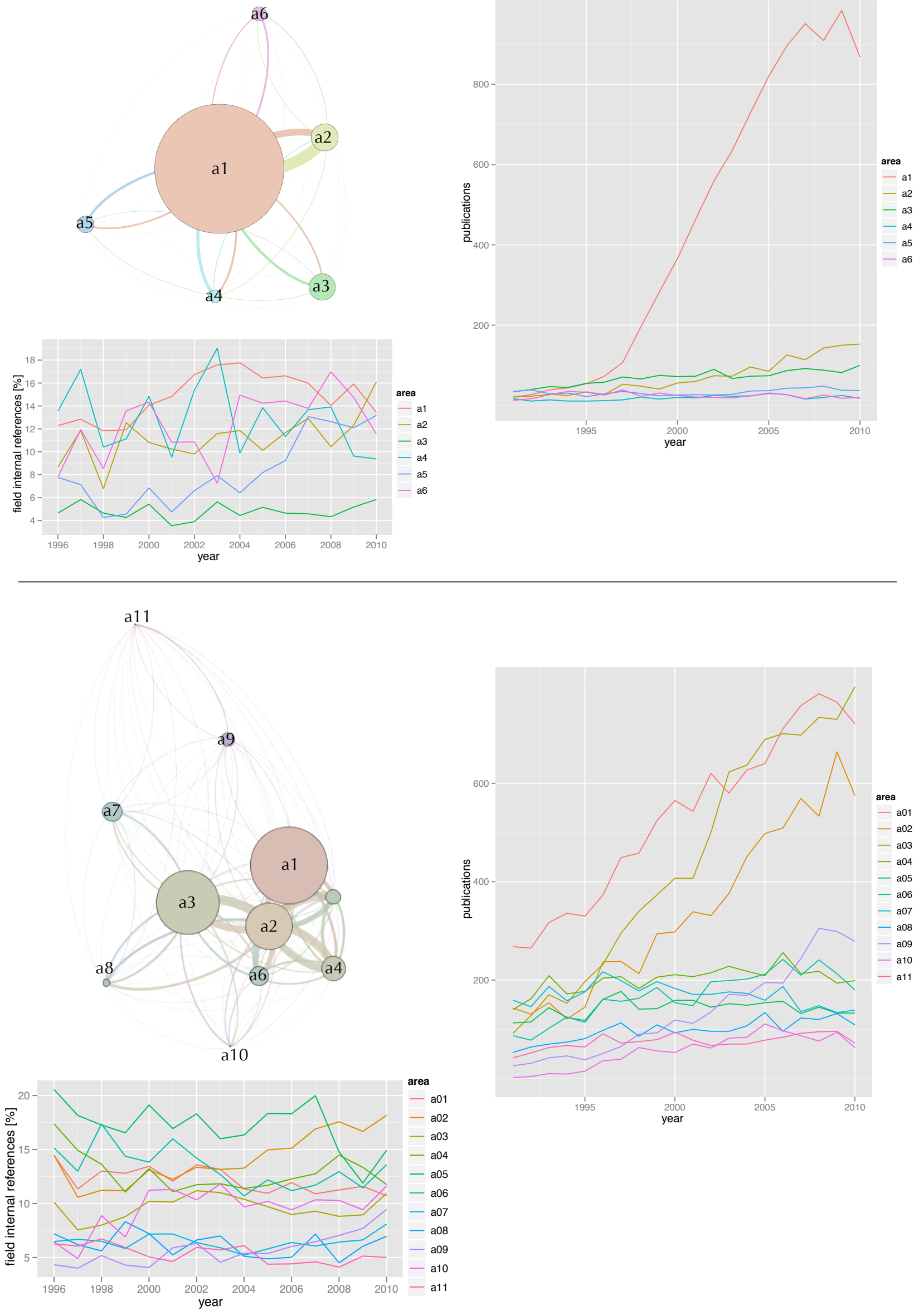

**Figure 6: Citation network of the 6 largest topic areas in field 1 (top), and 11 largest topic areas in field 2 (bottom). Nodes represent document clusters extracted from document citation networks for field 1 and field 2. Node size is determined by number of documents in a topic area, link strength by the number of citations between the documents in the respective pair of topic areas, and link color indicates the source node of the directed links. Network visualization done with gephi using the Force Atlas 2 algorithm. Below the network visualizations, the evolution of reference inclusion rates between 1996 and 2010 for field 1 (top) and field 2 (bottom) is shown, and to the right, the evolution of area sizes over the twenty-year time frame from 1991 to 2010 for field 1 (top) and field 2 (bottom).**

an explanation for this relative preference for polymer topic areas for citations from documents in area 1. The smaller areas seem to fall into two distinct sets:

**[Set 1]** *2:Polymer Chemistry*, *4:Polymer Chemistry/Organometallics*, *6:Polymer Chemistry/Molecular Catalysis*

**[Set 2]** *3:Organometallics & Applications in Inorganic Chemistry*, *5:Development of supported catalysts*

In field 2, area *2:Surface Science* stands out as a central knowledge resource in the field receiving disproportionately many citations from six other topic areas. Overall the network has an elongated structure of overlapping groupings of areas. One can may perceive of it as two sets of areas at the extreme ends of the structure, that connect in the middle through areas 2,3, and 8.

**[Set 1]** *1:Physical Chemistry & Chemical Physics*, *4:Physical Chemistry & Chemical Physics*, *5:Dynamic Processes*, *6:Applied Surface Science*, *10:Applied Physics*

**[Set 2]** *7:Inorganic Chemistry*, *9:Inorganic Chemistry*, and *11:Bio-Inorganic Chemistry*.

**[Set 3 (connecting)]** *2:Surface Science*, *3:Materials Science*, and *8:Catalysis*

#### 5.2.4 Social Connectedness of Topic Areas

In this section we explore the preference that authors who publish in one area have for publishing also in other areas. We suggest that this kind of association between two topic areas implies a form of social connectedness as those authors interact with their colleagues e.g. at conferences and workshops or through peer review, and thereby facilitate the informal flow of knowledge and social mobility between topic areas. The resulting association heat maps and affinity networks for field 1 and field 2 are depicted in figure 7.

We observe for field 1 a relative isolation of area *1:Catalyst Development and Organic Chemistry*. Also, authors in area *5:Development of supported catalysts* show a particularly strong preference for contributing also to area *4:Polymer Chemistry/Organometallics*[15]. Overall the global structure of topic area relationships depicted in the affinity network shows two sets of areas, one oriented towards polymer chemistry, and another oriented towards organometallics and inorganic chemistry. They are strongly coupled via the two polymer chemistry areas 4 and 6:

**[Set 1]** *2:Polymer Chemistry*, *4:Polymer Chemistry/Organometallics*, *6:Polymer Chemistry/Molecular Catalysis*

**[Set 2]** *3:Organometallics & Applications in Inorganic Chemistry*, *5:Development of supported catalysts, 4:Polymer Chemistry/Organometallics*, *6:Polymer Chemistry/Molecular Catalysis*

For field 2 we obtain a more dense affinity network compared to the citation based affinity network. We find especially strong mutual affinities between areas *9:Inorganic Chemistry*, and *11:Bio-Inorganic Chemistry*, as well as between areas *7:Inorganic chemistry* and *9:Inorganic chemistry* and between.areas *4:Physical Chemistry & Chemical Physics* and *5:Dynamic Processes*, and between areas *2:Surface Science* and *5:Dynamic Processes* and between *6:Applied Surface Science* and *10:Applied Physics.* The network shows a similar pattern to the citation network, however suggesting a slightly different subdivision of areas, namely into three overlapping sets:

**[Set 1]** *1:Physical Chemistry & Chemical Physics*, *2:Surface Science*, *4:Physical Chemistry & Chemical Physics*, *5:Dynamic Processes*, *6:Applied Surface Science*

**[Set 2]** *2:Surface Science*, *3:Materials Science*, *6:Applied Surface Science*, *8:Catalysis*, and *10:Applied Physics*

**[Set 3]** *7:Inorganic chemistry*, *8:Catalysis*, *9:Inorganic Chemistry*, and *11:Bio-Inorganic Chemistry*.

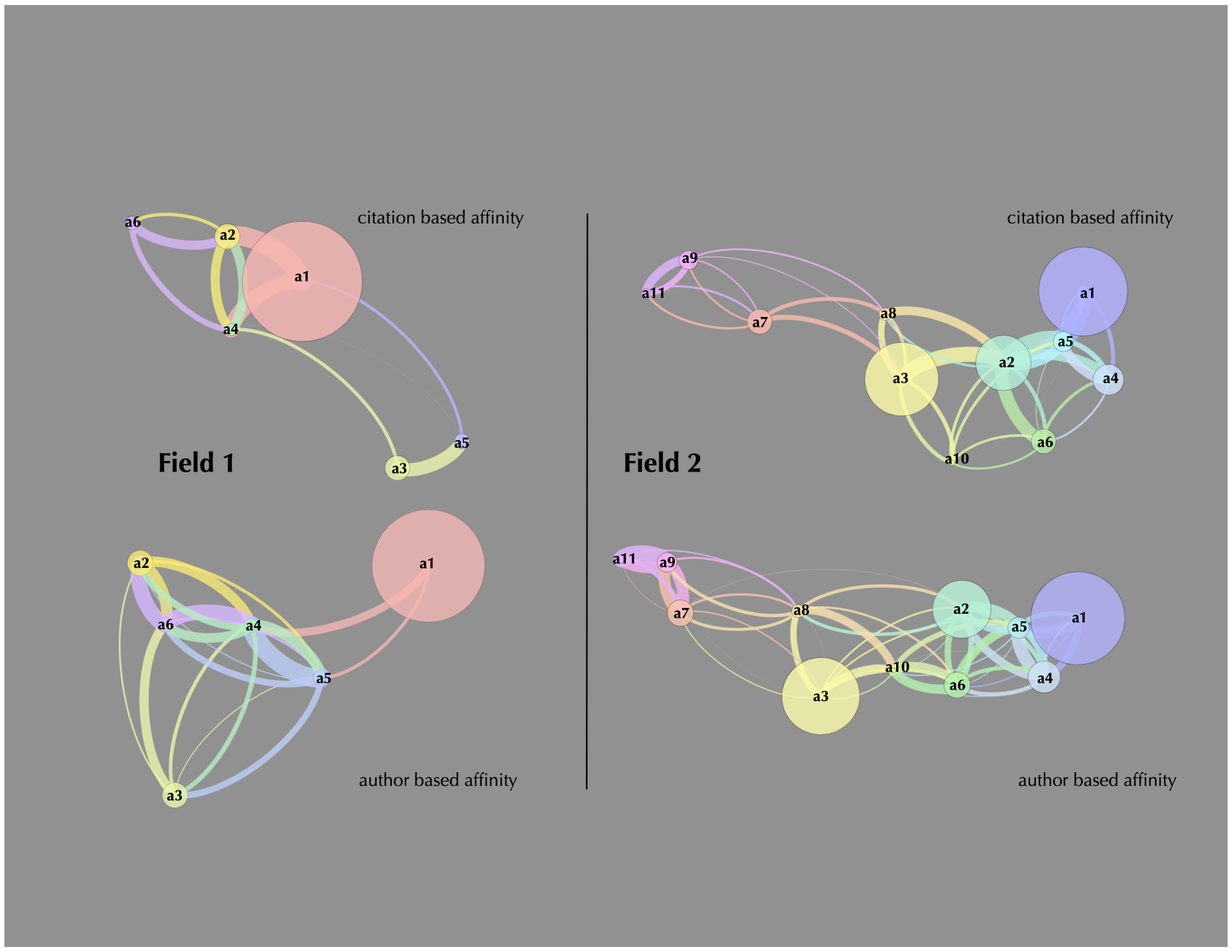

**Figure 7: Affinity diagrams for citation based association (top) and author activity based association (bottom) between topic areas in field 1 (left) and in field 2 (right). The affinity networks highlight positive associations between areas. Link colors indicate the source node of the directed links. The lay-out was produced using gephi and the Force Atlas 2 algorithm. Data tables with the exact values are provided in the supplementary material.**

### 5.2 Mapping Community Structures as Overlays on Collaboration Network

The basic structural differences between the group collaboration networks in field 1 and field 2 have already been discussed in Velden et al. (2010), albeit for networks generated from non-disambiguated data. Those structural features persist after author name disambiguation, as can be seen in figure 8. Specifically these are the fragmentation of the group collaboration network in field 1 versus the integration of co-author clusters in field 2 into one large network component; and the lower rate of group collaboration in field 1 versus field 2. In field 1 only 9.8% of clusters in the giant component of the co-author network have one or more collaborative relationships to other clusters, whereas in field 2 this number is more than twice as high, namely 22.9%. In the remainder of the section we create geographic and topical overlay maps by projecting information about the co-author clusters onto the group collaboration network to visualize community structures.

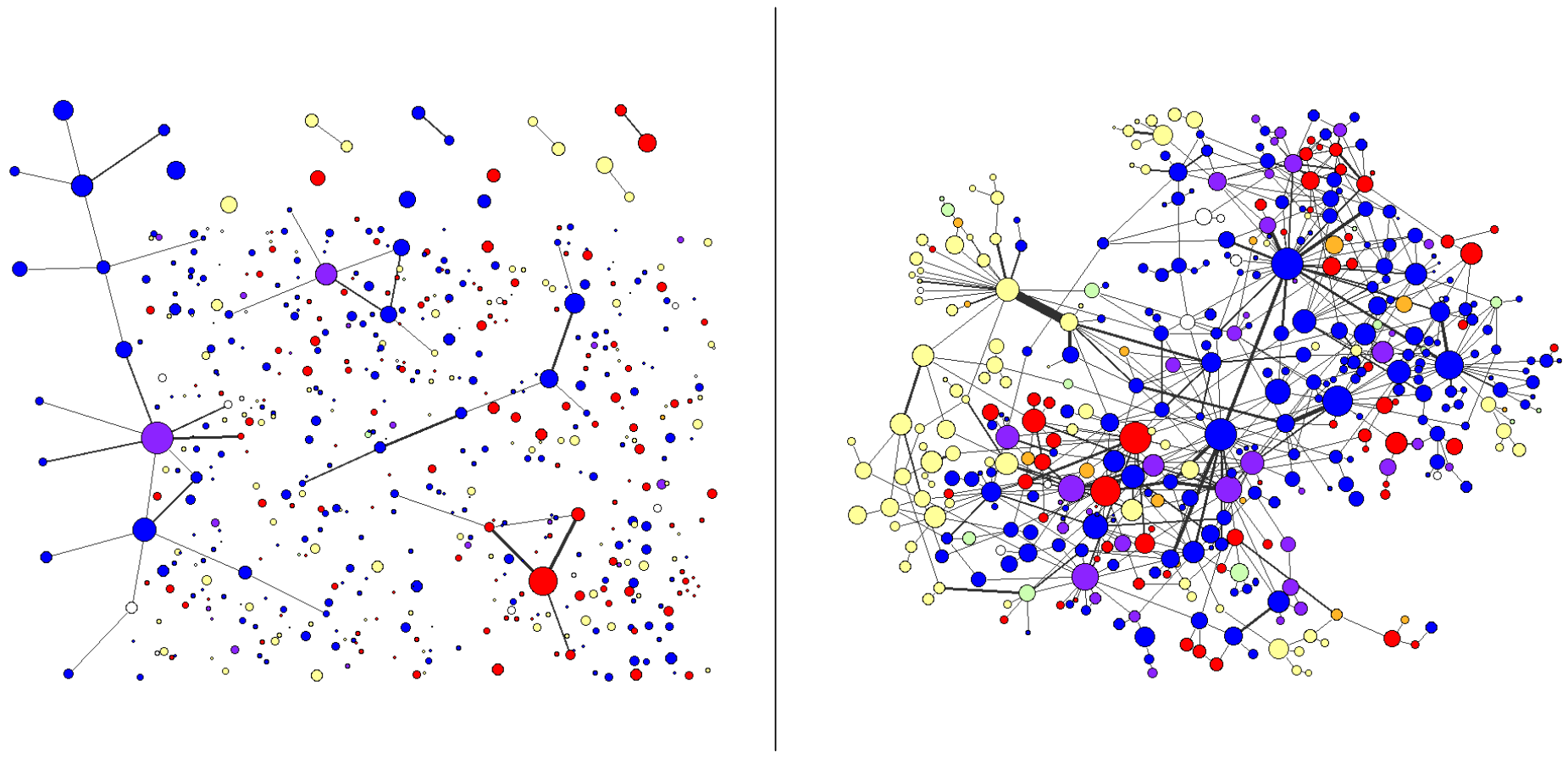

**Figure 8: Inter-group collaboration network build from the giant component of the co-author network in each field. For field 1 (left) all co-author clusters, with and without group collaboration links are shown. For field 2 (right) only the giant component of the group collaboration network is shown, omitting 118 co-author clusters with no or few collaboration links to other co-author clusters. Node sizes represent number of co-authors in a cluster and link strength the number of co-author relationships between authors in the respective clusters. Node color indicates the dominant geographical affiliation of a co-author cluster (light yellow=Asia, blue=Europe, red=North America, light green=Asia-Europe, violet=Europe- North America, white=Other).**

The geographic overlay maps in figure 8 depict the dominant geographic affiliations for each co-author cluster in the group collaboration network (see section 3.2.5.). For both fields we observe geographical ordering in the group collaboration networks. A quantitative evaluation shows a preference for collaborations between groups with specific geographic affiliations (details and data are provided in supplementary material). The geographical ordering is most pronounced in field 1, but it is present also in field 2. For both fields we find that Asian groups and North American groups each show a strong preference for collaborating with groups from their own continent. European groups, on the other hand, show neutral propensity to collaborate with other European groups. In field 1, European groups show a strong preference for collaboration with mixed European/North American groups, and in field 2, European groups show a preference for collaborating with Asian/European groups and European/North American groups.

Inspecting the geographic overlay maps in figure 8 we observe that in field 2 the overlay map highlights the internal clustering of Asian groups in two larger and three smaller clusters while there is a more cohesive interlinking of European groups with North American and mixed European/North American groups. The map suggests an integrative role of European/North American affiliated groups, which represent 10% of the co-author clusters in the group collaboration network. Indeed, the European/North American affiliated groups in the network have on average disproportionally many collaboration links (3.4), versus European groups (2.0), North American groups (1.9), or Asian groups (1.4), along with a preference for interlinking with European groups, North American groups, and other European/North American groups (data included in supplementary material).

### 5.2.2 Topical Overlay

The topic overlay maps show for each major topic area in a field the varying intensities with which groups are publishing in the respective area (see section 3.2.5.). We make the following observations:

The topic overlay map in figure 9 highlights that most of the groups in field 1 are active in the topic area *1:Catalyst Development and Organic Chemistry*. Collaborating groups often have their activity focuses in the same topic areas, but occasionally groups collaborate that have their activity focus in different fields. However, the collaboration network is too small and fragmented for a meaningful comparison to the topic area affinity structure presented in the previous section.

The topic overlay map in figure 10 echoes the topic area substructure revealed for field 2 with the help of the affinity networks. Co-author clusters with high activity in topic area *2:Surface Science* bridge two disparate regions of the group collaboration network: one region in the upper right side that includes on the upper left the majority of co-author clusters with high activity in topic areas *7:Inorganic Chemistry*, *9:Inorganic Chemistry*, and *11:Bio-Inorganic Chemistry* and to the right the majority of co-author clusters with high activity in topic areas *3:Materials Science*, *6:Applied Surface Science*, *8:Catalysis*, and *10:Applied Physics*. On the opposite, lower left side of the collaboration network is a region that includes the majority of co-author clusters with high activity in topic areas *1:Physical Chemistry & Chemical Physics*, *4:Physical Chemistry & Chemical Physics*, and *5:Dynamic Processes*. There are deviations from this overall pattern however, in the sense that there is some ’fuzziness’ in the mapping of some of the topic areas onto the collaboration network. Some co-author clusters that are very active in an area are located at opposite ends of the collaboration network; see e.g. areas 2, 3, 6, and 7. Upon inspection of the geographically labeled collaboration network in figure 8, this fuzziness could be explained with geography as a competing ordering principle in the collaboration network that is particular strong for Asian research groups in the field.

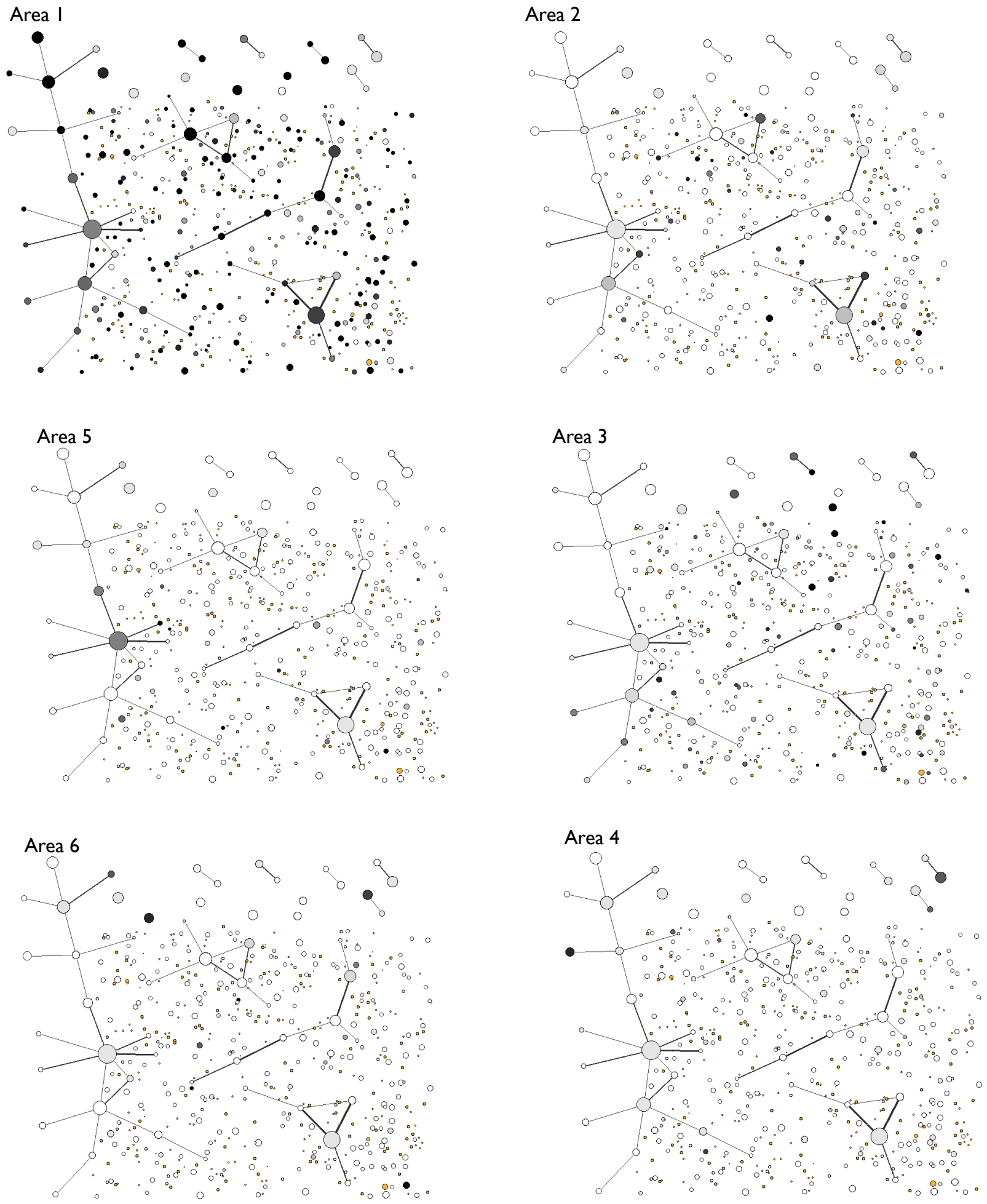

| Field 1 | topic area description | most popular journals (# of articles) | Field 1 | topic area description | most popular journals (# of articles) |
|---|---|---|---|---|---|
| **Area 1** (9024 articles) | Catalyst Development and Applications in Organic Chemistry | TETRAHEDRON LETT (881)<br>ORGANIC LETT (810)<br>J OF ORGANIC CHEMISTRY (684)<br>JACS (641)<br>ORGANOMETALLICS (508) | **Area 5** (664 articles) | Catalyst development (supported catalysts) | JACS (53)<br>J MOLECULAR CATALYSIS A-CHEMICAL (53)<br>ORGANOMETALLICS (50)<br>APPL. CATALYSIS A-GENERAL (33)<br>J CATALYSIS (27) |
| **Area 2** (1430 articles) | Polymer Chemistry (synthesis of block copolymers) | MACROMOLECULES (352)<br>J POLYMER SCIENCE PART A (116)<br>JACS (81)<br>MACROMOLECULAR CHEM &PHYSICS (58)<br>POLYMER (55) | **Area 6** (492 articles) | Polymer Chemistry (molecular catalysis, synthetic metals) | MACROMOLECULES (53)<br>J POLYMER SCIENCE PART A-POLYMER CHEM. (47)<br>J MOLECULAR CATALYSIS A-CHEMICAL (34)<br>J ORGANOMETALLIC CHEMISTRY (29)<br>POLYMER (25) |
| **Area 3** (1389 articles) | Organometallics (applications in inorganic chemistry) | ORGANOMETALLICS (383)<br>JACS (148)<br>INORGANIC CHEMISTRY (144)<br>J ORGANOMETALLIC CHEMISTRY (113)<br>DALTON TRANSACTIONS (112) | **Area 4** (374 articles) | Polymer Chemistry/Organo-metallics | MACROMOLECULES (58)<br>ORGANOMETALLICS (34)<br>J ORGANOMETALLIC CHEMISTRY (26)<br>J MOLECULAR CATALYSIS A-CHEMICAL (26)<br>J POLYMER SCIENCE PART A-POLYMER CHEM. (20) |

**Figure 9: Topic overlay on collaboration network of field 1. Nodes represent co-author clusters, links intergroup collaboration. Node color indicates relative intensity of publication activity of a cluster in the topic area ranging from white (= 0%) to black (> 90%).**

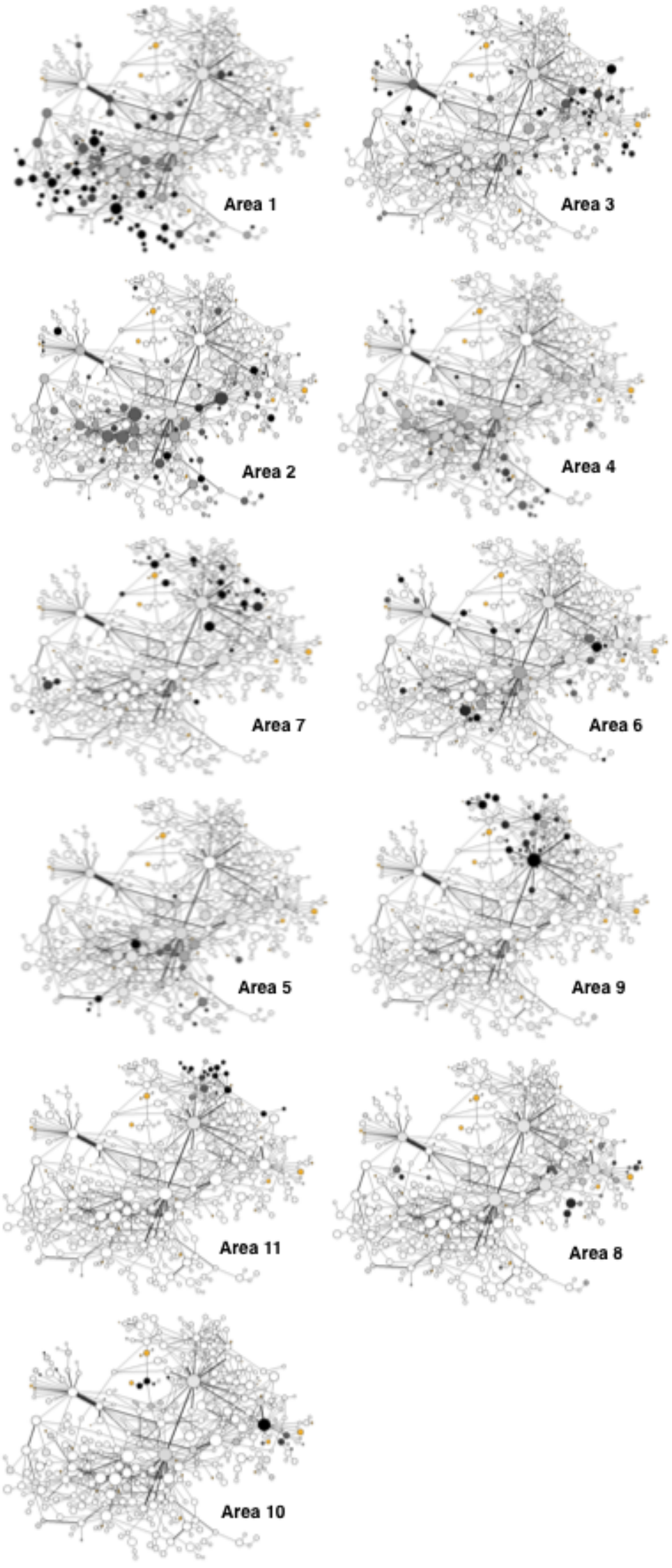

| Field 2 | topic area description | most popular journals (# of articles) |
|---|---|---|
| Area 1 (10632 articles) | Physical Chemistry Chemical Physics | J OF CHEMICAL PHYSICS (2435)<br>J OF PHYSICAL CHEMISTRY A (1262)<br>CHEMICAL PHYSICS LETTERS (864)<br>PHYSICAL CHEMISTRY CHEMICAL PHYSICS (390)<br>J OF PHYSICAL CHEMISTRY B (277) |
| Area 3 (8902 articles) | Materials Science | J OF PHYSICAL CHEMISTRY B (503)<br>LANGMUIR (416)<br>JACS (363)<br>CHEMISTRY OF MATERIALS (299)<br>J OF PHYSICAL CHEMISTRY C (264) |
| Area 2 (6819 articles) | Surface Science | J OF CHEMICAL PHYSICS (860)<br>PHYSICAL REVIEW B (535)<br>SURFACE SCIENCE (367)<br>J OF PHYSICAL CHEMISTRY A (361)<br>CHEMICAL PHYSICS LETTERS (340) |
| Area 4 (4028 articles) | Physical Chemistry Chemical Physics | J OF CHEMICAL PHYSICS (512)<br>CHEMICAL PHYSICS LETTERS (319)<br>J OF PHYSICAL CHEMISTRY A (244)<br>PHYSICAL REVIEW B (237)<br>J OF MOLECULAR STRUCTURE-THEOCHEM (95) |
| Area 7 (3394 articles) | Inorganic Chemistry | INORGANIC CHEMISTRY (464)<br>ORGANOMETALLICS (442)<br>J OF ORGANOMETALLIC CHEMISTRY (391)<br>J CHEM SOC - DALTON TRANSACTIONS (191)<br>INORGANICA CHIMICA ACTA (172) |
| Area 6 (3370 articles) | Applied Surface Science | NUCL. INSTR. & METH. IN PHYS RES SECT. B (341)<br>PHYSICAL REVIEW B (253)<br>SURFACE SCIENCE (196)<br>APPLIED SURFACE SCIENCE (136)<br>J OF APPLIED PHYSICS (121) |
| Area 5 (2853 articles) | Dynamic Processes | J OF CHEMICAL PHYSICS (250)<br>PHYSICAL REVIEW A (227)<br>PHYSICAL REVIEW B (216)<br>Z. F. PHYS. D-ATOMS MOL. & CLUSTERS (201)<br>PHYSICAL REVIEW LETTERS (179) |
| Area 9 (2704 articles) | Inorganic Chemistry | INORGANIC CHEMISTRY (345)<br>JACS (157)<br>PHYSICAL REVIEW B (117)<br>DALTON TRANSACTIONS (96)<br>INORGANICA CHIMICA ACTA (93) |
| Area 8 (1954 articles) | Catalysis | J OF CATALYSIS (199)<br>J OF PHYSICAL CHEMISTRY B (136)<br>APPLIED CATALYSIS A-GENERAL (98)<br>CATALYSIS LETTERS (81)<br>J OF PHYSICAL CHEMISTRY (76) |
| Area 11 (1501 articles | Bio-Inorganic chemistry | BIOCHEMISTRY (228)<br>J OF BIOLOGICAL CHEMISTRY (141)<br>JACS (124)<br>INORGANIC CHEMISTRY (95)<br>J OF BIOLOGICAL INORGANIC CHEMISTRY (64) |
| Area 10 (1113 articles) | Applied Physics | NUCL. INSTR. & METH. IN PHYS RES SECT. B (141)<br>PHYSICAL REVIEW B (83)<br>APPLIED PHYSICS LETTERS (75)<br>J OF APPLIED PHYSICS (67)<br>J OF NON-CRYSTALLINE SOLIDS (48) |

**Figure 10: Topic overlay on collaboration network of field 2. Nodes represent co-author clusters, links intergroup collaboration. Node color indicates relative intensity of publication activity of a cluster in the topic area ranging from white (= 0%) to black (> 90%).**

## 6 Discussion

Below we discuss the results, focusing on field delineation (section 6.1), on the extraction of community structures of research specialties (section 6.2), and on the utility of this for scaling-up ethnographies (section 6.3).

### 6.1 Field Delineation

Depending on the field, defining a lexical query that captures relevant publications with high accuracy is challenging. One of the two fields in our study is defined by a single, standardized term that is very reliably used in publications in this field. Hence, we consider the data set of field 1 as a rather accurate representation of the publication output in this field (to the extent that it is captured by the WoS database). Indeed, we find that the growth curve of annual publication numbers and the topic area-specific growth curves match the popular historical account of the field and its evolution, in particular the onset of rapid growth in the mid 1990's and the dynamics of growth driven by the topic area within the field, which focuses on the design of catalysts and their application in organic chemistry.

In field 2 the temporal evolution of the field is dominated by growth in four major topic areas located across the entire area affinity network. So far our ethnographic field study provides only a preliminary account of field growth enabled by the technical advancement of expensive, shared experimental facilities that research groups in this field have access to. Overall field 2 grew less dynamically and especially the growth curve of area 1 shows saturation suggesting a mature field. We would expect to be able to identify additional sub-community specific factors that shape the growth of the field in future research led by insights into the community structures in this field.

We have put forward the suggestion that the comparison of reference inclusion rates between topic areas may be helpful, to identify potential areas of overlap with other fields. The fact that in both fields most topic areas with low reference inclusion rates are related to inorganic chemistry may point to a discipline specific referencing pattern (e.g. higher rates of references to older literature or application of specific methods developed in another field) rather than indicating overlap with another closely related field. On the other hand, if we look at the topic overlay on the group collaboration network for field 2 we observe that those topic areas with lowest reference inclusion rates are mapped to the outer regions of the group collaboration network, whereas the topic areas with higher reference inclusion rates are mapped onto the center of the network. This would suggest that lower reference inclusion rates do indicate relative marginality of a topic area within the field. At this point, we regard the evidence for the interpretation of variations in reference inclusion rates as inconclusive.

Finally, the methods we have developed to optimize precision and recall of the lexical query, are heuristic in character. The method for optimizing recall relies on subjective decisions whether a cluster of publications in the self-citation network of a renown researcher should be considered part of the targeted field or not, and whether the set of individuals selected represents the field with sufficient breadth. Similarly, the decision whether to exclude certain terms in the lexical query to increase precision is ambiguous. However, in contrast to approaches to field delineation that require the ad-hoc setting of numerical cut-off parameters (e.g. Zitt & Bassecoulard 2006), the decision is supported by the ability to trace back and link the antagonism between topic areas in a data set to those topic areas' topical focus and hence to terms included in the lexical query. Hence the field delineation is guided by the social topology of the field that emerges from researchers' choices to cross field boundaries with their research and publication activities. A mixed method approach further supports this process by providing an understanding of actors' distinctions with regard to conceptual definitions of research topics developed in the ethnographic field study.

### 6.2 Community Structures Within Research specialties

The topic areas extracted from the two research specialties by clustering of the citation network reflect (sub)disciplinary orientations, along with specific research foci. This multi-disciplinary composition of each of the two fields corroborates insights from our ethnographic studies (from participant accounts, observations and scrutiny of conference programs) that scientists from a range of different (sub)disciplinary backgrounds[16] are active in the same research specialty.

The analysis of topical relatedness and social connectivity between topic areas reveals pronounced patterns of affinity and antagonism. More clearly than a mere citation network (as depicted in figure 6) the affinity networks suggest topical and social closeness and distance between topic areas within a field. These patterns vary depending on whether we consider inter-citation between areas or author activity across areas, however the global structures are very similar. They point to sub-communities within research specialties that reflect differences in disciplinary orientation and, we suggest in research practices. We found that some of the antagonisms detected between topic areas matches with sentiments expressed by our field study participants. For example members in an experimental physics group in field 2 that is mostly active in areas 1 and 5 commented on their collaboration with a physical chemistry group whose work is focused in area 3. They expressed reservations about the kind of research pursued by the other group and invoked a 'dirty chemistry' trope referring to chemical synthesis and ill-defined research objects in contrast to the pure, well-defined objects ('every atom counts') that they themselves, as physicists, preferred to study.

Finally, for a field such as field 2, where a substantial proportion of groups is part of the collaboration network, the topic overlay map provides a compelling insight into the ordering of the group collaboration network. The dominant feature of the topic overlay map for field 2 is the ordering of the group collaboration network first by topic areas, and then by topic area affinities. The distribution of co-author clusters in the collaboration network as depicted in figure 10 mirrors the global structure of the affinity networks obtained from citations and author activity. This implies that groups with a research focus in an area tend to collaborate with groups that have their focus of activity in the same area or a related, positively associated area.

For some topic areas we find that geographic ordering of the group collaboration network competes with this topical ordering. Taking into account that data about the growth of field 2 shows a recent dramatic increase in share of publication output in particular by researchers in China, we hypothesize that this competing geographic ordering may indicate a yet incomplete integration of these research groups into the global science system (Veugelers 2009).

The affinity networks together with the topical overlay map suggests a subdivision of field 2 into three smaller sub-communities and it could be argued that they represent research specialties in their own right. However, these sub-communities are all but disjoint. Almost 50% of groups and the large majority of authors are joined together in the giant component of the co-author network. This implies at least overlap between those sub-communities in the form of either direct collaboration between groups in different sub-communities or individuals participating in multiple communities either concurrently or over time. Further, as evidenced by the topical overlay map as well as ethnographic observations, groups collaborate across even distant topic areas, conferences in the field bring these different sub-communities together, and field study participants acknowledge the similarity of research questions pursued by research across these communities. Hence, we suggest reserving the term research specialty for this largest collective of researchers that collectively produce new knowledge in a domain, and refer to these smaller entities as topic areas (if defined based on the citation network of publications) or sub-communities (when referring to the researchers involved).

### 6.3 Scaling-up of Ethnographies

Ethnographic field research is expensive in terms of time resources. Instead of relying on a statistically representative sample, ethnography relies on the strategic selection of field sites. Network analysis of the relationships between actors in the domain of interest can support this selection process (Howard 2002). We suggest that the community structures revealed in this analysis relate to differences in disciplinary orientation and research practices, underlining that fields at the specialty level cannot be assumed to be homogenous in this regard. Therefore, network analysis cannot only highlight particular actors of interest, but through the mapping of community structures as demonstrated in this article, support the strategic selection of field sites for study and comparison.

To zoom in on one of the sub-communities revealed in our network analysis will usually reduce the diversity of (sub)disciplinary research practices of the research groups included. However, to assume in turn that these sub-communities are largely homogeneous in their research practices would be gravely erroneous. Instead the analytic unit we are offered is that of a sub-community that can be distinguished in its (sub)disciplinary orientation from other sub-communities within the field, but whose core of closely communicating research groups still includes a range of (sub)disciplinary backgrounds that come along with distinct research practices. For example, from our field studies we know that the sub-community in field 2 that is composed of research groups active in topic areas 1,4, and 5 brings together groups with backgrounds in physical chemistry, experimental atomic and molecular physics, and theoretical physics. This finding echoes observations in an exploratory study of the extent and limitations of cross-disciplinarity within a research specialty studying aspects of molecular motors in bio-nano technology (Rafols & Meyer 2006). To the extent we could determine, this sub-community does not further subdivide into sub-communities that would be more homogeneous with regard to disciplinary backgrounds and research practices. This seems to be either due to lack of critical mass (there are just too few theoretical physics groups working on this type of research questions to constitute a community) or because the interaction of those groups with complementary skills is essential to produce results that are considered as legitimate and meaningful (such as the need to combine theoretical models with empirical data).

This implies that the ethnographer is still faced with a complex unit of analysis to study and develop a deeper understanding of. However we suggest that the kind of community analysis introduced here supports better informed choices of field sites supporting the selection of units of analysis that correlate with a closely interacting community, and the design of comparative studies that may throw distinct differences into relief. The scaling-up of ethnographic observations is a critical step for comparative science studies that aim for a systematic understanding of the co-construction of research practices and the social organization of scientific communities. Behaviors such as openness and sharing of knowledge between research groups are shaped by the competition dynamics within a field that in turn are shaped by a number of research specific epistemic and material factors (Velden 2011a). The advantage and relevance of having network analytic tools that help identify sub-community structures is that we can control for epistemic and social differences introduced by sub-community structures.

## 7 Conclusions

The scientific community that collectively produces the knowledge base of a research field is an important unit of analysis for understanding how epistemic and social factors influence e.g. communication behaviors, but it is challenging to capture. We demonstrate here, that publication networks can be 'mined' as part of a mixed method approach to reveal and map the complexities of a scientific community. We argue that insights into the community structures within a research field are crucial for the strategic selection of ethnographic field sites. By better understanding community structures of scientific fields and hence the context of local ethnographic observations, we aim to increase the confidence about the domain of validity of ethnographic results, and hence their value for systematic comparisons across scientific fields.

Future work will be directed at adding temporal resolution of community structures for improved guidance in the selection of field sites and to track the evolution of community structures over time. Also, based on strategic sampling of field study sites supported by insights into community structures, we plan to investigate variations of openness and sharing behaviors within fields to systematically study the influence of epistemic factors.

## Acknowledgements

We gratefully acknowledge the participants in our field studies. Without their support and their consent to participate in the study this work could not have been accomplished. This material is based upon work supported by the National Science Foundation through a Doctoral Thesis Improvement Grant No. 0924445 and under Grant No OCI-1025679.

**Appendix: Explanation of Terms Used**

**research field / research specialty:** by field we here refer to a research field at the level of a research specialty. Whereas the terms discipline (such as ’chemistry’ or ’physics’) or sub-discipline (such as ’organic chemistry’) refer to larger, bureaucratized units of teaching and academic employment, research specialty refers to the knowledge base generated collectively by a self-organizing loosely coordinated community of researchers (Gläser 2006). The scientific community of researchers that is active in a research specialty can have a variety of sizes, from a few researchers to hundreds or perhaps even thousands of researchers (depending on the size of the smallest collective unit needed to produce new knowledge claims). As discussed in the article, research specialties (as a cognitive concept) overlap, as do the communities of scientists contributing to the knowledge bases of research specialties.

**scientific community:** the collective of researchers active in a research specialty. Membership in a scientific community is based on self-perception and expressed in research activities oriented towards contributing to the shared knowledge base of that research specialty (Gläser 2006).

**(institutional) research group:** a research group is an organizational unit at a university or research institute, usually of collocated researchers led by a principal investigator (PI). Typically, the PI is the only ‘constant’ presence in the group over time. In our experience, research groups oftentimes actively contribute to several research specialties. Since not all group members of an institutional research group are doing research in the same research field, a co-author cluster extracted from a field specific data set will capture only a partial view of an institutional research group. Instead in the kind of sciences that we have been studying so far - characterized by a median of around 4 co-authors per publication - the co-author clusters that we extract with Rosvall’s algorithm come closest to capturing functional research groups (defined below).

**functional research group:** to be distinguished from a research group as organizational unit of an academic institution is the collective of researchers that Seglen & Asknes (2000) have called ‘functional research groups’. These typically are research collectives including a collocated PI-led group of researchers extended by closely cooperating domestic or international colleagues or visiting scientists. Most co-author clusters extracted in our study will not correspond to institutional research groups, but represent functional research groups. For lack of temporal resolution the co-author clusters in our study are not a direct representation of a functional research group as instantiated at any particular point in time, but represent its accumulated bibliographic footprint over a 20-year time frame.

**principal investigator (PI):** a term borrowed from US American funding agency parlance. By this term we refer to the senior leader of an institutional research group or a functional research group; in an institutional academic setting this is typically a researcher at a professorial rank.

[1] Terms in bold, italicized fonts are explained in the appendix.
[2] See explanation of term in appendix.
[3] We are hesitant however to adopt the term ‘network ethnography’ as it suggests a misleading identification of the research object in the ethnography, scientific communities, with networks. Empirical evidence suggests that scientific communities may contain networks, but are not identical with and constituted by networks (Gläser 2006, p. 28-30).
[4] For a review of this newly emerging field see e.g. Börner et al. (2007).
[5] Given they were continuously active over the entire time period, and not just small because of a temporally limited participation in the field.
[6] See explanation of term in appendix.
[7] See explanation of term in appendix.

[8] The clustering of citation networks to study research specialties has been pioneered by Small (1973) based on a subset of highly cited documents and using co-citation instead of direct citations. Recently, Shibata et al. (2009) conducted a comparison of direct citation, co-citation, and bibliographic coupling to detect research fronts, and concluded that direct citation outperformed the other two approaches in detecting research fronts as it identified large and emerging clusters earlier. In addition, they observed that direct citation networks had the largest clustering coefficients, an indication of higher content similarity. At the time we developed our approach this finding suggested that direct citation is a good choice for revealing the topical substructure in a research specialty. Since then a very comprehensive analysis of a large bio-medical corpus by Boyack and Klavans (2010) has delivered conflicting results, concluding that bibliographic coupling and co-citation analysis outperform direct citation in identifying research fronts. In future work it may be of interest to assess the difference that the exact construction of the network to be clustered makes for the extraction of topic areas and subsequent community structure analysis.

[9] Since our focus here is on teasing out the interrelatedness of areas, we do not include the source area as one of the potential target areas. Hence, we here disregard self-referentiality although it may represent a potentially interesting property of topic areas.

[10] In the fields studied here, these hub nodes can be typically identified as the PIs of research groups and represent the research continuity of those research groups. If we included all publications co-authored by any member of a co-author cluster we would introduce noise in the form of publications that cluster members made while they were members of other groups.

[11] Homonymy occurs when multiple authors with the same last name and initials get merged into a single author entity.

[12] This subject filtering is used to exclude irrelevant document sets retrieved from other fields of research because of alternate meaning of terms used in the query.

[13] infomap_undir, available from the homepage of Martin Rosvall (http://www.tp.umu.se/~rosvall/code.html).

[14] An alternative way for topic area identification would be the cognitive mapping approach that has been recently proposed by Rafols et al. (2010) that projects publication data onto the science map constructed from inter-citation of ISI WoS subject categories. However, these subject categories are one step removed from the journal titles as they classify and group journals. Most journal titles are fairly descriptive, so we here relied on a direct interpretation of these journal titles rather than the derived subject categorization.

[15] A plausible link given that polymers are one of the popular supports for catalysts in field 1.

[16] We identify disciplinary background with a scientist's formal training and nominal faculty appointment.

Supplementary Material to

**The Extraction of Community Structures from Publication Networks to Support Ethnographic Observations of Field Differences in Scientific Communication**

Theresa Velden*, Carl Lagoze

tvelden@umich.edu, clagoze@umich.edu

## Part 1: Field growth

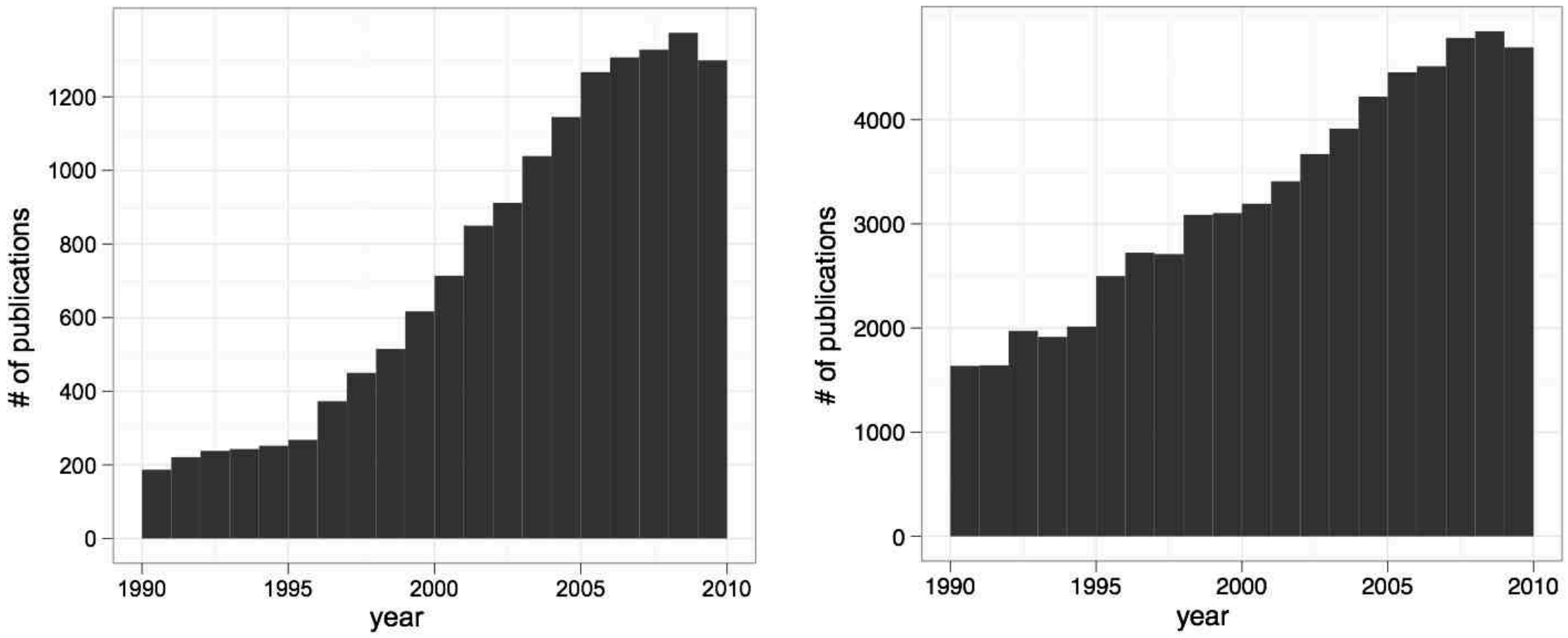

**Figure 2: Annual output of publications in field 1 (left), and field 2 (right)**

## Part 2: Data tables of associations underlying the affinity networks in figure 7

Data table 1: Field 1 citation association between topic areas (ordered by topic areas size)

| | Target areas | | | | | |
|---|---|---|---|---|---|---|
| **Source areas** | a1 | a2 | a3 | a5 | a6 | a4 |
| a1 | 0 | 16.3 | -19.2 | -2.5 | -4.1 | 13.2 |
| a2 | 6.8 | 0 | -15.7 | -10.6 | 2.8 | 7.7 |
| a3 | -0.9 | -5.8 | 0 | 11.1 | -1.9 | 2.9 |
| a5 | 2.6 | -7.4 | 2.1 | 0 | -2.4 | 0.2 |
| a6 | -0.5 | 7.7 | -6 | -4.3 | 0 | 4.9 |
| a4 | -0.6 | 8.1 | -5.9 | -2.8 | 1.8 | 0 |

Data table 2: Field 1 author activity association between topic areas (ordered by topic areas size)

| | **Target areas** | | | | | |
|---|---|---|---|---|---|---|
| **Source areas** | a1 | a2 | a3 | a5 | a6 | a4 |
| a1 | 0.0 | -3.43 | -3.6 | 3.91 | -0.46 | 8.98 |
| a2 | -4.44 | 0.0 | -4.77 | 3.61 | 12.81 | 11.52 |
| a3 | -4.41 | 2.02 | 0.0 | 1.25 | 10.34 | 4.17 |
| a5 | -9.39 | 1.41 | 7.15 | 0.0 | 7.45 | 21.01 |
| a6 | -9.51 | 17.24 | -4.04 | 2.96 | 0.0 | 16.81 |
| a4 | -10.99 | 7.78 | 5.34 | 12.22 | 10.64 | 0.0 |

Data table 3: Field 2 citation association between topic areas (ordered by topic areas size)

| | **Target areas** | | | | | | | | | | |
|---|---|---|---|---|---|---|---|---|---|---|---|
| **Source areas** | **a1** | **a3** | **a2** | **a4** | **a7** | **a6** | **a5** | **a9** | **a8** | **a11** | **a10** |
| **a1** | 0 | -28.3 | 32.5 | 14.2 | -16.9 | -7.8 | 42.9 | -8.3 | -11.9 | -11.9 | -10.8 |
| **a3** | -32.2 | 0 | 51.4 | -12.2 | -4.8 | -0.8 | -1.1 | -14.6 | 21.4 | -10.9 | 14 |
| **a2** | -25.3 | -8.7 | 0 | 36.2 | -21.6 | 13.4 | 58.1 | -19.5 | 8.2 | -19.4 | -2.7 |
| **a4** | -15 | -20.9 | 46.9 | 0 | -15.2 | 6.9 | 33.5 | -12.2 | -12.1 | -4.7 | -9.1 |
| **a7** | -13.8 | 16.9 | -0.8 | -4.9 | 0 | -9.1 | -6.5 | 6.4 | 13.7 | 8.1 | -5.1 |
| **a6** | -17.7 | -6.3 | 44.1 | 10.9 | -16.1 | 0 | 3.5 | -14.4 | -6 | -10.7 | 7.8 |
| **a5** | 4.9 | -20.6 | 37.1 | 17.4 | -15.8 | -2.1 | 0 | -14.1 | -12.2 | -11 | 0.5 |
| **a9** | -1.3 | 3.5 | -4.3 | -5.9 | 5 | -6.5 | -6.1 | 0 | 6.1 | 20.7 | -3.6 |
| **a8** | -14.4 | 11.2 | 29.9 | -10.3 | -2.7 | -5.5 | -8.4 | -3.4 | 0 | -4.2 | -3.8 |
| **a11** | -7.4 | -4.1 | -6.1 | -2.4 | 7.6 | -4.7 | -4.1 | 31.6 | 6.5 | 0 | -2.7 |
| **a10** | -15 | 13.6 | 10.3 | -8 | -8.5 | 8.6 | 10.2 | -7.2 | -0.9 | -5.5 | 0 |

Data table 4: Field 2 author activity association between topic areas (ordered by topic areas size)

| | **Target areas** | | | | | | | | | | |
|---|---|---|---|---|---|---|---|---|---|---|---|
| **Source areas** | **a1** | **a3** | **a2** | **a4** | **a7** | **a6** | **a5** | **a9** | **a8** | **a11** | **a10** |
| **a1** | 0 | -27.87 | 12.67 | 24.2 | -13.35 | 4.84 | 41.51 | -7.17 | -9.34 | -13.84 | -10.55 |
| **a3** | -24.12 | 0 | 6.05 | -4.13 | 3.72 | 11.76 | 5.66 | 0.26 | 20.59 | -15.32 | 21.52 |
| **a2** | -7.77 | -11.41 | 0 | 14.38 | -13.23 | 17.69 | 33.37 | -20.1 | 11.36 | -16.2 | 2.94 |
| **a4** | 7.76 | -28.73 | 26.4 | 0 | -19.64 | 13.85 | 42.45 | -20.43 | -13.68 | -16.64 | 3.47 |
| **a7** | -13.44 | 4.74 | 1.24 | -3.79 | 0 | -11.88 | -6.96 | 37.17 | 7.9 | 1.97 | -6.6 |
| **a6** | -14.19 | -11.15 | 20.66 | 13.44 | -19.69 | 0 | 29.63 | -14.27 | -5.09 | -13.42 | 30.22 |
| **a5** | 16.52 | -25.21 | 27.34 | 25.02 | -21.12 | 3.04 | 0 | -20.58 | -13.56 | -15.29 | 4.36 |
| **a9** | -12.9 | -5.01 | -8.13 | -5.27 | 37.41 | -9.41 | -7.92 | 0 | 7.51 | 35.78 | -3.58 |
| **a8** | -21.61 | 0.73 | 10.74 | -8.05 | 9.99 | 7.31 | -9.24 | 11.19 | 0 | -5.2 | 26.7 |
| **a11** | -12.76 | -6.39 | -9.55 | -4.65 | 14.8 | -6.79 | -6.86 | 54.89 | 4.95 | 0 | -5.19 |
| **a10** | -17.87 | 4.39 | -1.76 | 0.8 | -11.56 | 29.77 | 17.71 | -2.21 | -1.28 | -7.89 | 0 |

## Part 3: Country Specific Collaboration Preferences

Preferences are expressed as deviation in percent from a null model that assumes that groups randomly chose groups with other continent affiliations to collaborate with, such that the proportion of collaboration links groups with a certain continent affiliation have with other groups depends on the relative number of groups for each continent affiliation. (AS: Asia, EU: Europe, NA: North America)

Data Table 5: Field 1 Geographical Propensity for Group Collaboration

| | **Preference for Collaboration Partners From [relative deviation from null model]** | | | | | |
|---|---|---|---|---|---|---|
| **Continent affiliation** (average node degree*) | AS | EU | NA | AS/EU | AS/NA | EU/NA |
| AS (0.6) | **414%** | -60% | -100% | N.A. | N.A. | -100% |
| EU (1.1) | -79% | -5% | -79% | N.A. | N.A. | **575%** |
| NA (1.0) | -100% | -77% | **390%** | N.A. | N.A. | 243% |
| AS/EU (0) | N.A. | N.A. | N.A. | N.A. | N.A. | N.A. |
| AS/NA (0) | N.A. | N.A. | N.A. | N.A. | N.A. | N.A. |
| EU/NA (5.5) | -100% | 31% | -38% | N.A. | N.A. | -100% |

* Collaboration links to other groups in the (unweighted) collaboration network

Data Table 6: Field 2 Geographical Propensity for Group Collaboration

| | **Preference for Collaboration Partners From [relative deviation from null model]** | | | | | |
|---|---|---|---|---|---|---|
| **Continent affiliation** (average node degree*) | AS | EU | NA | AS/EU | AS/NA | EU/NA |
| AS (1.4) | **150%** | -56% | -30% | 30% | 90% | -45% |
| EU (2.0) | -70% | 22% | -44% | 86% | -31%. | **122%** |
| NA (1.9) | -50% | -41% | 115% | -21% | 113% | **122%** |
| AS/EU (2.5) | -30% | **49%** | -41% | 11% | -10%. | -66% |
| AS/NA (2.1) | 23% | -34% | **92%** | 8% | -100%. | 31% |
| EU/NA (3.4) | -78% | **33%** | 25% | -75% | -18% | 29% |

* Collaboration links to other groups in the (unweighted) collaboration network